\documentclass[onecolumn,twocolappendix,dvipsnames]{aastex63}

\usepackage{amsmath,amsbsy}
\usepackage{xcolor}
\usepackage[normalem]{ulem}

\newcommand{\rvec}{{\mathbf r}}

\newcommand{\bvec}{{\mathbf B}}
\newcommand{\Bvec}{{\mathbf B}} 
\newcommand{\jvec}{{\mathbf J}}
\newcommand{\Jvec}{{\mathbf J}}
\newcommand{\kvec}{{\mathbf k}}

\newcommand{\avec}{{\mathbf A}}
\newcommand{\nablavec}{\,\vec{\nabla}}

\newcommand{\hatz}{\,\hat{\mathbf z}}

\newcommand{\emfac}{\mu_0}
\newcommand{\jt}{\mathcal{T}}
\newcommand{\jp}{\mathcal{P}}
\newcommand*\Bell{\ensuremath{\boldsymbol\ell}}

\newcommand{\be}{\begin{equation}}
\newcommand{\ee}{\end{equation}}
\newcommand{\bea}{\begin{eqnarray}}
\newcommand{\eea}{\end{eqnarray}}
\newcommand{\beax}{\begin{eqnarray*}}
\newcommand{\eeax}{\end{eqnarray*}}
\newcommand{\ba}{\begin{array}}
\newcommand{\ea}{\end{array}}
\newcommand{\bed}{\begin{description}}
\newcommand{\ed}{\end{description}}
\newcommand{\blc}{\begin{list}{$\circ$}{}}
\newcommand{\blb}{\begin{list}{$\bullet$}{}}
\newcommand{\el}{\end{list}}
\newcommand{\ben}{\begin{enumerate}}
\newcommand{\een}{\end{enumerate}}

\def\laprox{\mathrel{\hbox{\rlap{\hbox{\lower4pt\hbox{$\sim$}}}\hbox{$<$}}}}
\def\gaprox{\mathrel{\hbox{\rlap{\hbox{\lower4pt\hbox{$\sim$}}}\hbox{$>$}}}}

\begin{document}

\title{The Photospheric Imprints of Coronal Electric Currents: \\
I. Magnetic Structure Near Polarity Inversion Lines}

\author{Brian T. Welsch}
\affil{Natural \& Applied Sciences,
University of Wisconsin-Green Bay, 2420 Nicolet Drive, Green Bay, WI 54311, USA}

\author{Peter W. Schuck}
\affil{Heliophysics Science Division, NASA Goddard Space Flight Center, 8800 Greenbelt Rd., Greenbelt, MD 20771, USA}

\author{Mark G. Linton}
\affil{Space Science Division, US Naval Research Laboratory, 4555 Overlook Ave, SW, Washington, DC, USA}

\begin{abstract}
Flares and coronal mass ejections (CMEs) are powered by magnetic energy stored in coronal electric currents.
Here, we use photospheric vector magnetic field observations to study currents in active regions 10930 and 11158, which both produced eruptive, X-class flares. 
We employ Gauss's separation method in Cartesian geometry to partition the photospheric field into three distinct components: (i) the toroidal field, $\bvec_T$, from vertical currents, $J_z$, passing through the photosphere; (ii) $\bvec^<$, from 
currents, $\Jvec^<$, flowing below it; 
and (iii) $\bvec^>$, from 
currents, $\Jvec^>$, flowing above it.
We refer to $\bvec^>$ as the photospheric imprint of coronal currents.
We give two representations of $\bvec^<$ and $\bvec^>$: (i) as second-order derivatives of poloidal potentials $P^<$ and $P^>$, respectively; and (ii) in terms of gradients of scalar potentials, with $\bvec^< = -\nablavec \psi^<$ and $\bvec^> = -\nablavec \psi^>$. 
The central polarity inversion line (PIL) in each region possesses magnetic structure similar to that reported in AR 12673 by Schuck et al. (2022):
(i) $\Bvec_T$, which arises from $J_z$, produces sheared fields along the PIL;
(ii) $\bvec^>$ exhibits large-scale, spatially coherent structure, consistent with $\Jvec^>$ flowing horizontally above and along the PIL; and
(iii) the near-PIL, horizontal currents $\Jvec_h^<$ and $\Jvec_h^>$ are roughly parallel. 
Because parallel currents attract, such parallel-current configurations are likely more stable than misaligned-current configurations.
Horizontal current $\Jvec_h^>$ flowing along a PIL increases $|\nablavec B_z|$ across the PIL, providing a physical explanation for previously reported empirical associations of strong, cross-PIL gradients in $B_z$ with flare and CME occurrence.
\end{abstract}

\section{Introduction}
\label{sec:intro}

Despite decades of study, the physical processes that generate solar
flares and coronal mass ejections (CMEs) are not fully understood.  A
core concept, referred to as the storage-and-release paradigm, is that
these events are powered by the release of magnetic energy stored in
coronal electric currents; see, e.g., the discussion of ``storage
models'' by \citet{Forbes2000}.  Accordingly, electric currents
flowing through the solar corona play key roles in determining when
and where these magnetically-driven events occur.  Indeed, if a
coronal magnetic field
lacked currents (and was thus curl-free by Amp\`ere's law, $\mu_0
\jvec = \nablavec \times \bvec$), then we expect that field would not produce flares or CMEs.
%
Because a curl-free vector field can be represented as the gradient of
a scalar potential, a current-free magnetic field is often referred to
as ``potential,'' and current-carrying fields are often called
``non-potential.''
%
%

The presence of currents in observed photospheric, chromospheric, and
coronal fields has been inferred from significant deviations between 
observed fields and potential-field models.  
While each particular choice of boundary conditions (BCs; i.e., Neumann, Dirichlet, or mixed) on a domain's boundaries uniquely determines a potential field within that domain, differing choices of BCs generally yield differing potential-field models, so selecting which potential model to compare with observations involves subjective choices.
In solar physics, potential fields have typically been derived using a Neumann boundary condition based on the inferred radial photospheric magnetic field (e.g., \citealt{Schatten1969, Altschuler1969}) from magnetograms (i.e., 2D maps of one or more magnetic field components), with differing assumptions about other boundaries.
\citet{Schrijver2005} compared observed coronal extreme ultraviolet (EUV) loops with field lines in potential models of the coronal field in several dozen active regions (ARs), and found flares to be significantly more likely in non-potential ARs.
In AR magnetograms, non-potential photospheric fields are
frequently observed near polarity inversion lines (PILs, where the
vertical component of the photospheric magnetic field changes sign):
near-PIL fields are often sheared (e.g., \citealt{Moore1987}),
meaning that they possess a strong component that runs {\em along} the PIL,
rather than {\em across} it as generic potential fields would.
Large flares and eruptions invariably start in the atmosphere
immediately above strongly sheared PILs.
Filaments and helmet streamers, which can erupt as CMEs, also exhibit
non-potential magnetic structures that imply the presence of electric
currents, such as field-aligned H-$\alpha$ fibrils in filaments whose
orientations are inconsistent with potential-field orientations (e.g.,
\citealt{Martin1998}), and above-limb rings of enhanced linear
polarization consistent with flux ropes viewed in cross section (e.g.,
\citealt{Dove2011}).

Since currents play key roles in determining when, where, and how
magnetic energy is released in flares and CMEs, understanding these
events requires addressing a fundamental question: What is the
structure of coronal electric currents?  Unfortunately, observations
provide only limited information about $\bvec$ in the corona —-- e.g.,
the line-of-sight-integrated transverse component of $\bvec$ above
the limb (e.g., \citealt{Lin2004, Tomczyk2008, Dove2011}), or field
strength $|\bvec|$ from radio gyrosynchrotron observations (e.g.,
\citealt{Brosius2006}). (We remark that the Frequency-Agile Solar Radiotelescope [FASR, \citealt{Gary2023}], if built, should greatly improve observational characterization of coronal magnetic fields.) Hence, the distribution of electric currents
in the corona ($\jvec$ as a function of space) cannot be directly
measured with present capabilities.  

Lacking direct measurements of the coronal vector magnetic field, modeling of the coronal field, by upward extrapolation from photospheric magnetograms, is often employed to 
the study of the structure of coronal currents.  One common approach
for such modeling is to assume that the
Lorentz force in the coronal field vanishes, $\jvec \times \bvec = 0.$ For
nonzero currents, this implies
%
\be \nablavec \times \bvec = \alpha \bvec ~, 
\label{eqn:fff} \ee
where, in the general case, $\alpha$ varies in space transverse to the field (but is constant along it). When $\alpha$ is
a function of position, solutions of this equation are referred to as
non-linear, force-free fields (NLFFFs).  Several methods for finding
NLFFFs have been developed; for a review, see, e.g.,
\citet{Wiegelmann2012}.  
It should be noted that the force-free assumption underlying this approach is, in general, incompatible with the non-force-free nature of photospheric fields (e.g., \citealt{Metcalf1995}).  
Discrepancies between observed photospheric fields and fields on the bottom boundaries of NLFFF models can arise from both observational errors in observed fields and simplifications in the models (perhaps chiefly, the assumption that $\jvec \times \bvec = 0$).   
\citet{DeRosa2015} noted that model-data discrepancies are often large and ``in excess of nominal
uncertainties in the data.'' 

A recent, exciting development in the study of solar magnetic fields 
promises progress in understanding coronal currents: the 
first-of-its-kind application of Gauss's
separation method \citep{Gauss1839} --- a technique with a long heritage in modeling the magnetic fields of Earth (e.g., \citealt{Schmidt1899, Langel1985}) and other solar system bodies (e.g., \citealt{Kivelson2002}) --- to photospheric vector magnetograms of solar active regions by \citet{Schuck2022}. 
For reviews of Gauss's method, see, for instance, \citet{Backus1986,Olsen2010}. 
\citet{Schuck2022} applied Gauss's method, which they call CICCI --- Carl's Indirect Coronal Current Imager --- to both synthetic magnetograms from MHD sunspot simulations and observed magnetograms of AR 12673, to separate the photospheric field into contributions produced by currents that flow: (i) through the surface; (ii) below the surface, in the convection zone; and (iii) above the surface, in the chromosphere/corona.  
They used the MHD simulation results to validate CICCI.  They then applied CICCI\footnote{A repository of CICCI and related routines is online at https://git.smce.nasa.gov/cicci} to magnetograms from  
the Helioseismic and Magnetic Imager (HMI; \citealt{Scherrer2012,schou2012}) aboard the {\em Solar Dynamics Observatory} (SDO; \citealt{Pesnell2012}) 
and found evidence for large-scale currents flowing above and along PILs in AR 12673. 
%

%
More recently, \citet{Titov2025} presented an alternative approach, which they call the magnetogram-matching Biot-Savart law (MBSL) method, for decomposing the photospheric field to isolate the contribution of above-surface currents to the horizontal photospheric field. The MBSL method is not inconsistent with Gauss's approach: the field components determined via MBSL are linear combinations of
the field components derived from Gauss's method.
Because the MBSL method combines field components that arise from separate source-current domains, its decomposition contains less information about the photospheric field's origins than Gauss's decomposition.

\par

\citet{Schuck2022} applied Gauss's method in Cartesian geometry to their MHD validation data, but did not explicitly describe their Cartesian version (derivable in the $R_\odot \to \infty$ limit).   
We discuss Gauss's separation method in Cartesian geometry in detail in Section \ref{sec:decomp} below, but one key strength of the method warrants emphasis at the outset of our explication: 
part of the observed
magnetic field at the photosphere can be {\em
  unambiguously} attributed to currents flowing above the photosphere.
Mathematically, in Cartesian coordinates with a flat photosphere at $z=0$, 
the observed photospheric field $\bvec^{\rm ph}(x,y)$ can be uniquely partitioned 
into four components,
\be \bvec^{\rm ph}(x,y)
= \bvec_T(x,y,0) + \bvec^<(x,y,0) + \bvec^>(x,y,0) + \bvec_M ~,
\label{eqn:b_decomp} \ee
where: $\bvec_T(x,y,0)$, the toroidal component of the
horizontal photospheric field, is due to currents, $J_z(x,y,0)$, that pass through
the photosphere; $\bvec^<(x,y,z)$ is the field in $z \ge 0$ due to currents $\jvec^<$ below the photosphere; $\bvec^>(x,y,z)$ is the field in $z \le 0$ due to currents $\jvec^>$ above it; and $\bvec_M$ is any mean field present at the photosphere.
For simplicity, we call all currents flowing above the photosphere 
``coronal,'' whether they flow in the chromosphere, transition region, 
or corona, and
we refer to $\bvec^>(x,y,0)$ (or $\bvec^<(x,y,0)$) as the {\bf photospheric imprint} of coronal currents (or interior currents, respectively).
While many models of coronal magnetic fields have assumed that coronal
currents only affect the horizontal photospheric field, Gauss's method reveals 
that these currents also produce a significant part of the observed
$B_z^{\rm ph}(x,y)$ (\citealt{Schuck2022}, and below).
We remark that Gauss's separation method makes no assumption about 
photospheric fields being force-free.

This paper uses Gauss's method to explore manifestations of near-surface coronal currents in photospheric magnetograms.  
Our results follow on the recent study by \citet{Schuck2022}, 
who took the key step of applying Gauss's separation
algorithm 
to solar fields measured by 
HMI.
Because this approach is relatively new within solar physics, we begin by
discussing the decomposition in Equation (\ref{eqn:b_decomp}) in the next Section.  
While \citet{Schuck2022} presented the decomposition in spherical coordinates, we present the decomposition in Cartesian coordinates. 
(We remark that, formally, a Cartesian result can be obtained by taking the $R_\odot \to \infty$ limit of the spherical version).
In addition to presenting Gauss's decomposition in the context of a poloidal-toroidal decomposition (PTD; see, e.g., \citealt{Chandrasekhar1961,Backus1986,Lantz1999, Fisher2010, Berger2018}; for earlier uses of this decomposition, see references in \citealt{Backus1986} below his equation [2]), as was done by \citet{Schuck2022}, we also present the decomposition in terms of a scalar potential for current-free fields. 
Then, in Sections \ref{sec:10930} and \ref{sec:11158}, we apply the method to decompose the photospheric magnetic fields of two active regions that produced X-class flares and CMEs: NOAA AR 10930, observed by the Hinode satellite, and AR 11185, observed by HMI. 
%
In Section \ref{sec:conclusions}, we discuss our key findings, their
possible applications, and directions for future work. 

\section{Decompositions of Photospheric $\bvec$}
\label{sec:decomp}

\citet{Gauss1839} presented a method to partition the magnetic field  on a spherical surface into two components, each with a distinct  source: either currents interior to the surface, or exterior to it. \citealt{Glassmeier2014} give a modern translation. 
Gauss was considering the properties of the Earth's surficial magnetic field, so assumed that no currents flowed {\em across} the surface. 
\citet{Backus1986} presented a generalization of Gauss's approach to cases when currents cross the surface, enabling the decomposition shown in Equation (\ref{eqn:b_decomp}). 
To incorporate currents through the surface of interest, a poloidal-toroidal decomposition of the surface magnetic vector field has been employed with Gauss's separation method (e.g., \citealt{Backus1986,  Olsen2010, Schuck2022}).  

\subsection{Poloidal - Toroidal Decomposition (PTD)}
\label{subsec:ptd}

Because our analysis will focus on structures within ARs that are relatively  small compared to the radius of the Sun, we will perform the separation in Cartesian coordinates.  In this subsection, we briefly review the PTD representation of magnetic fields, and in the next subsection, we outline Gauss's separation method using the Cartesian version of PTD. %
Because scalar-potential solutions to Laplace’s equation are commonly employed in modeling current-free coronal magnetic fields, in the subsection after the next one, we describe the separation in terms of such potentials.

Because any valid magnetic field $\bvec$ is divergence free, it can be expressed in terms of two scalar potentials, $T$ and $P$, that produce its toroidal and poloidal parts,
\be \bvec(x,y,z) =  \bvec_T(x,y,z) + \bvec_P(x,y,z) +  \bvec_M
~, \label{eqn:tor_pol_sum} \ee
where 
\bea \bvec_T &=& \nablavec \times (T(x,y,z) \hatz ) 
\label{eqn:tor_defn} \\
\bvec_P(x,y,z) &=&  \nablavec \times (\nablavec \times P(x,y,z) \hatz) \label{eqn:pol_defn} \\
&=&  \nablavec_h ( \partial_z P(x,y,z) ) 
- \nabla_h^2 P(x,y,z) \hatz  
~, \eea
where the surface gradient is $\nablavec_h \equiv (\partial_x, \partial_y, 0)^T$, and
the surface Laplacian operator is
$ \nabla_h^2 \equiv (\partial_x^2 + \partial_y)$. 
(It should be noted that toroidal and poloidal field components used here differ in meaning from the toroidal and poloidal coordinate directions often used for tori.)
Keeping with the notation of Equation (\ref{eqn:b_decomp}), $\bvec_M$ is any mean field that is present.
While our discussion focuses on the 2D, three-component photospheric field 
$\bvec^{\rm ph}(x,y) = \bvec(x,y,0)$, the PTD representation is valid for 3D fields.
The field's Cartesian components can be expressed in terms of the PTD potentials as
\bea 
B_x &=&  \partial_x ( \partial_z P ) + \partial_y T \label{eqn:ptd_bx} \\
B_y &=&  \partial_y ( \partial_z P ) - \partial_x T \label{eqn:ptd_by} \\
B_z &=& - \nabla_h^2 P  \label{eqn:ptd_bz} ~. \eea

Because both $\bvec_T$ and $\bvec_P$ are
derived from curls, the PTD representation of any 3D magnetic field is
divergence-free if consistent numerical schemes are used to compute
the curls of the potential functions and divergence of $\bvec$.  From
the structure of Equations (\ref{eqn:tor_defn}) and (\ref{eqn:pol_defn}), one can see that $\bvec$ 
can be expressed as the curl of a vector potential $\avec$ given by
\be \avec = (\nablavec \times P \hatz) + T \hatz + \nablavec\Lambda
\label{eqn:ptd_avec} 
~, \ee
where we have defined a gauge function $\Lambda$.

Equation (\ref{eqn:ptd_bz}) shows that the poloidal potential $P(x,y,z_i)$
in a layer at height $z_i$ can be found by solving a 2D Poisson equation with source $[-B_z(x,y,z_i)]$.
Amp\`ere's law can be used to solve for the toroidal potential $T(x,y,z_i)$ in that layer by taking the vertical component of the curl, or surface curl, of Equation (\ref{eqn:tor_pol_sum}), 
which yields another 2D Poisson equation,
\be \hatz \cdot (\nablavec \times \bvec)|_{z = z_i} = \mu_0 J_z(x,y,z_i) = 
-\nabla_h^2 T(x,y,z_i) ~. \label{eqn:jzi} \ee
(Because $\hatz \cdot \nablavec \times (\nablavec_h \partial_z P - \nabla_h^2 P \hatz) = 0$, $\bvec_P$ does not contribute here.) 
The horizontal divergence of the horizontal component of Equation
(\ref{eqn:tor_pol_sum}) is related to 
the vertical derivative of the poloidal potential, $\partial_z P$, by
yet another 2D Poisson equation,
\be \nabla_h^2 ( \partial_z P )|_{z = z_i}  
= (\nablavec_h \cdot \bvec_h)|_{z = z_i} 
\label{eqn:ptd_dpdz} ~.  \ee
(Note that $\nablavec_h \cdot (\nablavec_h \times T \hatz) = 0$, so $\bvec_T$ does not contribute here.) 
%
The vertical derivative of Equation (\ref{eqn:ptd_bz}) shows that $\partial_z B_z = - \nabla_h^2 (\partial_z P)$, ensuring the divergence-free condition on $\bvec$ is satisfied.   This relationship also illustrates the coupling of the field's 2D structure in a given layer to its 3D structure in the vicinity of that layer.


%
%
%
%

Readers should be aware that care must be taken when solving Equations (\ref{eqn:jzi}) and (\ref{eqn:ptd_dpdz}) for the $T$ and $\partial_z P$ potentials, respectively, on a finite domain. 
In a bounded region, general solutions to these 2D Poisson equations include solutions to their homogeneous versions (2D Laplace equations), i.e., surface-curl-free and horizontal-divergence-free fields.  The potentials for these ``source-free'' fields are harmonic functions, and their form depends upon the assumed BCs on the region. 
%
%
For example, because the horizontal poloidal field, $\nablavec_h (\partial_z P)$, arises from a gradient, choosing either periodic or homogeneous BCs for $\partial_z P$ is incompatible with a mean horizontal poloidal field.
Hence, if mean field components are present, then they must be explicitly added when using either homogeneous or periodic BCs
to ensure that 
\be \bvec_h(x,y,z_i) = \bvec_T(x,y,z_i) + \bvec_{P,h}(x,y,z_i) + \bvec_{M,h} ~. \label{eqn:bh_match} \ee
(Any mean vertical field, $B_{M,z}$, consistent with the presence of a monopole source, must be poloidal. Such a field could be included by either (i) adding $B_{M,z}$ directly to $\bvec$ or (ii) including an additional poloidal potential $P'$ that has a surface Laplacian that does not vary horizontally, $\nabla_h^2 P' = B_{M,z}$, and whose vertical derivative vanishes at the photosphere,
$\partial_z(\nabla_h^2 P')|_{z = 0} = 0$,  
for consistency with Equation (\ref{eqn:ptd_dpdz}).)
\citet{Fisher2010} included extensive discussion of the role of non-periodic BCs for PTD potentials on finite domains. \citet{Bhatia2014} recommends using the free-space Green's function solutions to the 2D Poisson's equations for $T$ and $\partial_z P$ 
(see, e.g., \citealt{Berger2018}), with any residual values $(\bvec^{\rm ph}_h - \bvec_T - \bvec_{P,h})$ on the domain perimeter (attributed to sources exterior to the domain) being matched by appropriate harmonic potentials. 

%

The key issue for any bounded patch is that the BCs applied on the patch's perimeter can change the vector fields associated with the field's $T$ and $\partial_z P$ components: it is possible to add equal and opposite horizontal magnetic fields that have both zero surface curl and zero horizontal divergence to $\bvec_T$ and $\bvec_{P,h}$ without violating Equation (\ref{eqn:bh_match}), so the decomposition of $\bvec_h$ into toroidal and poloidal parts is not unique.
Changes in this decomposition could affect inferences about currents at, above, and below the photosphere that produce the photospheric field. Practically speaking, though, for active region magnetograms with only weak, spatially incoherent fields along the domain perimeter --- that is, lots of quiet sun between the AR and the boundary --- the fields arising from gradients in harmonic potentials (which, as solutions to the 2D Laplace equation, take their extremal values on the boundaries) cannot significantly alter the decomposition of strong fields within ARs.  

We remark that ambiguities in harmonic potentials do not arise in the spherical harmonic approach of \cite{Gauss1839} and \cite{Schuck2022}, because the domain is simply connected (see discussion of related issues in \citealt{Schuck2026}).  
In practice, lack of full-sphere, ``4$\pi$'' magnetogram coverage of the Sun means that assumptions must be made about unobserved, far-side photospheric magnetic fields as well as poorly-observed near-limb fields on the front side. The assumed fields can introduce harmonic poloidal and toroidal components even in areas with well observed fields.   
In addition, the Sun's observed large-scale field typically exhibits a small but significant monopole moment (which can arise from instrumental effects as well as incomplete coverage), sometimes dealt with by subtracting off a small, spatially constant radial field (e.g., \citealt{DeRosa2012}). 

\subsection{Gauss's Separation in terms of PTD}
\label{subsec:ptd_separation}

In terms of the PTD formalism, we write the Cartesian version of the decomposition of the photospheric vector magnetic field in Equation (\ref{eqn:b_decomp}) as
\be
\bvec^{\rm ph}(x,y) =
\nablavec \times T(x,y,0) \hatz 
+ \nablavec_h \partial_z P^<(x,y,0) - \nabla_h^2 P^<(x,y,0) \hatz 
+ \nablavec_h \partial_z P^>(x,y,0) - \nabla_h^2 P^>(x,y,0) \hatz ~,
\label{eqn:b_decomp2}
\ee
where we have assumed the photospheric magnetic field is measured in the $z=0$ plane.   
%
We show in the Appendix that, under appropriate assumptions, $P$ is a function only of toroidal currents, and $T$ is a function only of poloidal currents. Then we show that $P$ can be expressed as the sum of a term which depends only on the toroidal currents in $z > 0$ and a term which depends only on the toroidal currents in $z < 0$. Thus, $\bvec_T$ is created by the poloidal currents in the system, $\bvec^>$ is generated by the toroidal currents above the photosphere, and $\bvec^<$ is generated by the toroidal currents below the photosphere.
The vertical current $J_z$ that flows across the photosphere is thus poloidal, and produces the toroidal magnetic field's scalar potential $T$, which is found by solving Equation (\ref{eqn:jzi}) at $z=0$ using the  observed $\bvec^{\rm ph}$.
%
Currents flowing below the photosphere produce $\bvec^<(x,y,0)$, equal to 
\be \bvec^<(x,y,0) = \nablavec_h \partial_z P^<(x,y,0) - \nabla_h^2 P^<(x,y,0) \hatz
~, \label{eqn:blt_ptd} \ee
while currents flowing above the photosphere produce $\bvec^>(x,y,0)$, equal to 
\be \bvec^>(x,y,0) = \nablavec_h \partial_z P^>(x,y,0) - \nabla_h^2 P^>(x,y,0) \hatz
~. \label{eqn:bgt_ptd} \ee
Thus, $\bvec^<(x,y,0)$ and $\bvec^>(x,y,0)$ are poloidal. While we could explicitly denote this with a $P$ subscript, we opt not to for notational simplicity.

\begin{figure}[ht]
\includegraphics[width=3.0in]{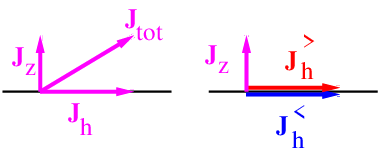}
\caption{\textsl{ Left: The total electric current, $       \jvec_{\rm tot}$ (tilted magenta vector), at a point     on the photosphere (thin black line) can be 
    decomposed into vertical $(J_z)$ and horizontal $(\jvec_h)$ components.  
    Right: We split any $\jvec_h$ at $z = 0$ into horitonal components flowing differentially above and below the photosphere. Currents at any height above or depth below the photosphere are denoted $\jvec^>$ and $\jvec^<$, respectively.  }}
\label{fig:split_jh}
\end{figure}

To partition the field in this way, we assume a zero-thickness photosphere, 
meaning that we formally partition any horizontal current density at the photosphere into components that flow either above or below the $z = 0$ plane, as illustrated in Figure \ref{fig:split_jh}.
Although magnetographs' spectral lines form over a finite height range (e.g., $\sim$300 km for HMI's Fe I 6173\AA; \citealt{Fleck2011}), this zero-thickness assumption is a good approximation for the structures that are our focus, which extend horizontally over a few Mm or more. 
(While the horizontal current density may be finite in the photosphere, the integral of it over the zero-thickness photosphere is zero, meaning that, in this approximation, the horizontal current density at the photosphere does not produce a magnetic signature at $z = 0$ [or above or below it].) 
Thus, only the vertical component of current flows ``at'' the photosphere.  
The treatment developed by \citet{Backus1986} does permit including a horizontal surface current (or magnetic tangential discontinuity) at the photosphere, with a $\delta$-function current density at $z = 0$. Such a current would produce a normal field at the photosphere. We assume, however, that such a singular current distribution is unphysical.  (\citet{Schuck2022} also exclude a singular surface current at the photosphere, per their footnote 5.)

The total poloidal field at the photosphere is the superposed field from potentials due to currents above and below the photosphere.    
Equation (\ref{eqn:ptd_dpdz}) evaluated at $z = 0$ implies
\be (\nablavec_h \cdot \bvec_h^{\rm ph}) 
= \nabla_h^2 \partial_z P^<(x,y,0) 
+ \nabla_h^2 \partial_z P^>(x,y,0) 
~. \label{eqn:ptd_dpdz_ltgt} \ee
%
(Since the horizontal divergence of a surface curl vanishes, the toroidal component
of $\bvec_h^{\rm ph}(x,y)$ does not contribute here.)
To relate the vertical photospheric field, $B_z^{\rm ph}(x,y)$, to its source currents, the differential form of Biot-Savart law \citep{Jackson1975} is relevant:
\be d\Bvec(\rvec)= \frac{\mu_0}{4 \pi}
\frac{\jvec(\rvec_s) \times (\rvec - \rvec_s)}{ |\rvec - \rvec_s|^3} dV
~, \label{eqn:bs_law} \ee
where $\rvec_s$ is the coordinate vector of a source point with an
infinitesimal current-carrying segment $[\jvec(\rvec_s) dV]$ that produces a magnetic field $d\bvec$ at the point with coordinate vector $\rvec$.
Each such current segment should be understood as a small portion of a larger current system.
From Equation (\ref{eqn:bs_law}), it is clear that 
the vertical current at the photosphere, $J_z(x,y)$,
produces no vertical field at the photosphere; hence, $B_z^{\rm ph}(x,y)$ is entirely due to currents above and below $z = 0$, and
%
%
\be B_z^{\rm ph}(x,y) 
= -\nabla_h^2 P^<(x,y,0) 
- \nabla_h^2 P^>(x,y,0)
\label{eqn:ptd_lapp_ltgt} ~. \ee
Note a key implication of Equation (\ref{eqn:ptd_lapp_ltgt}), which was emphasized by \citet{Schuck2022}: part of
the observed vertical field at the photosphere is due to coronal currents.

The next steps in the separation of $\bvec^{\rm ph}$ exploit the relationship of $P^<$ to $\partial_z P^<$ and of $P^>$ to $\partial_z P^>$ on $z = 0$.   
Recall that $P^<$ (respectively, $P^>$) is current-free for $z > 0$ (respectively, $z < 0$).
Per Backus (1986, his Equation [58c]), \nocite{Backus1986} the poloidal potential for a magnetic field that is current-free in a region obeys the 3D Laplace's equation in that region, meaning
\bea  
\nabla^2 P^< &=& 0, \quad z \in (0, +\infty) \label{eqn:laplace_poloidal_lt} \\
\nabla^2 P^> &=& 0, \quad z \in (-\infty, 0) \label{eqn:laplace_poloidal_gt} ~.
\eea
Here, we restrict our analysis of each poloidal potential to a region separate from its source currents;  Equations~(\ref{eqn:PLT})\--(\ref{eqn:PGT}) in the Appendix show the equations obeyed by each potential over the full domain. %
%
%
As is well known for solutions to Laplace's equation (see, e.g., \citealt{Jackson1975}), specifying a solution's value on the domain boundary (a Dirichlet BC) completely determines the solution's normal derivative on the boundary; conversely, specifying the solution's normal derivative on the boundary (a Neumann BC) completely determines the solution's value on the boundary. 
This implies that $P^<$ on $z=0$ is not independent of $\partial_z P^<$ on $z=0$, and, similarly, $P^>$ on $z=0$ is not independent of $\partial_z P^>$ on $z=0$. 
In fact, the two BCs in Equations (\ref{eqn:ptd_dpdz_ltgt}) and (\ref{eqn:ptd_lapp_ltgt}) provide two constraints that determine the two unknown solutions to Equations (\ref{eqn:laplace_poloidal_lt}) and (\ref{eqn:laplace_poloidal_gt}).
%


%


We first express the potentials $P^<$ and $P^>$ in terms of Fourier transforms as
\be P^{<(>)}(x,y,z) = \frac{1}{(2\pi)^2}
\int_{-\infty}^{+\infty} {\rm d}k_x \int_{-\infty}^{+\infty} {\rm d}k_y \,
\tilde P^{<(>)}(\kvec) \, {\rm e}^{{\rm i}k_xx + {\rm i}k_yy -(+) k_h z}
~, \label{eqn:p^pm_def} \ee
where the tilde denotes the 2D Fourier transform of a function, i.e.,
\be \tilde{f}(k_x,k_y) \equiv 
\int_{-\infty}^{+\infty} {\rm d}x \int_{-\infty}^{+\infty} {\rm d}y \,
f(x,y) \, {\rm e}^{-{\rm i}k_xx - {\rm i}k_yy}
~. \label{eqn:p_fourier} \ee
For each potential to solve the 3D Laplace equation, we require $k_h \equiv |\sqrt{k_x^2 + k_y^2}|$.  (Including the magnitude bars here is equivalent to always taking the positive root.)   Because $P^< \to 0$ as $z \to +\infty$, each of its Fourier modes decays as exp$(-k_h z)$, while $P^> \to 0$ as $z \to -\infty$, implying that each of its modes decays as exp$(+k_h z)$.

In terms of these Fourier transforms,  Equation (\ref{eqn:ptd_lapp_ltgt}) provides a joint Dirichlet BC for these functions,  
\bea B_z^{\rm ph}(x,y) 
&=& \frac{1}{(2\pi)^2} \int_{-\infty}^{+\infty} {\rm
  d}k_x \int_{-\infty}^{+\infty} {\rm d}k_y \, (\tilde P^<
(k_x,k_y) +\tilde P^> (k_x,k_y) ) \, k_h^2 \, e^{{\rm i}k_xx + {\rm
    i}k_yy} ~. \label{eqn:diri_ppm_source_fft} \eea
Denoting the Fourier transform of $B_z^{\rm ph}(x,y)$ as $\tilde B_z^{\rm ph}(k_x,k_y)$, we have
\be \tilde P^< (k_x,k_y) + \tilde P^> (k_x,k_y)
= \frac{
\int_{-\infty}^{+\infty} {\rm d}x
\int_{-\infty}^{+\infty} {\rm d}y \, B_z^{\rm ph}(x,y)
    {\rm e}^{-{\rm i}k_xx - {\rm i}k_yy} }{k_h^2}
= \frac{\tilde B_z^{\rm ph}(k_x,k_y)}
{k_h^2}
~. \label{eqn:dirichlet_ppm} \ee

In terms of the Fourier transforms of $P^<$ and $P^>$, Equation (\ref{eqn:ptd_dpdz_ltgt}) provides a joint Neumann condition for the two potentials.  Denoting $\tilde \bvec_h^{\rm ph}(k_x, k_y)$ as the 2D Fourier transform of $\bvec_h^{\rm ph}(x,y),$ we have 
\bea
\nablavec_h \cdot \bvec_h^{\rm ph}(x,y) 
&=&
\nablavec_h \cdot \left ( \frac{1}{(2\pi)^2} \int_{-\infty}^{+\infty} {\rm d}k_x
\int_{-\infty}^{+\infty} {\rm d}k_y \, 
\tilde{\bvec}_h^{\rm ph}(k_x,k_y) {\rm e}^{ {\rm i}k_xx + {\rm i}k_yy} \right ) 
  \\
  %
  &=& 
 \frac{1}{(2\pi)^2} \int_{-\infty}^{+\infty} {\rm d}k_x
\int_{-\infty}^{+\infty} {\rm d}k_y \, 
    {\rm i} \kvec_h \cdot \tilde{\bvec}_h^{\rm ph}(k_x,k_y)
    {\rm e}^{ {\rm i}k_xx + {\rm i}k_yy} \label{eqn:divhbh_fourier}
            \\
%
&=& \nabla_h^2 \partial_z \left . \left [
\frac{1}{(2\pi)^2} \int_{-\infty}^{+\infty} {\rm d}k_x \int_{-\infty}^{+\infty}
{\rm d}k_y \, [ \tilde P^<(k_x,k_y) {\rm e}^{-k_h z} + \tilde P^>(k_x,k_y) {\rm e}^{+k_h z} \,]  {\rm e}^{{\rm i}k_xx + {\rm i}k_yy } \right ] \right \vert_{z = 0} \\
\label{eqn:neumann_ppm1}
&=&
\frac{1}{(2\pi)^2} \int_{-\infty}^{+\infty} {\rm d}k_x \int_{-\infty}^{+\infty}
{\rm d}k_y \, k_h^3 \, [ \tilde P^<(k_x,k_y) - \tilde P^>(k_x,k_y) \, ]  {\rm e}^{{\rm i}k_xx + {\rm i}k_yy} 
~, \label{eqn:neumann_ppm2}
\eea
%
where the differing $z$-dependences of $P^<$ and $P^>$ imply that this condition involves a {\em difference} of their spectral functions.
Equations (\ref{eqn:divhbh_fourier}) and (\ref{eqn:neumann_ppm2}) imply that
\be \tilde P^<(k_x,k_y) - \tilde P^>(k_x,k_y) 
= \frac{{\rm i} \kvec_h \cdot \tilde{\bvec}_h^{\rm ph}(k_x,k_y)}{k_h^3}
~. \label{eqn:neumann_ppm3} \ee

While Equation
(\ref{eqn:neumann_ppm3}) involves a difference of $\tilde P^>$ and
$\tilde P^<$, Equation (\ref{eqn:dirichlet_ppm}) involves their {\em sum}.  
This is a key aspect of Gauss's separation method: the structure of
the horizontal and vertical components of the field in the separation
surface constrain the sum and the difference, respectively, of the
interior and exterior potentials' spectral functions.
(This same idea is emphasized by \citet{Olsen2010} for the spherical case.)
These differing dependencies thus provide two independent equations for two
unknown spectral functions, $\tilde P^<$ and $\tilde P^>$.

The sum and the difference of Equations (\ref{eqn:neumann_ppm3}) and
(\ref{eqn:dirichlet_ppm}) can then be used to find both spectral
functions, $\tilde P^<$ and $\tilde P^>$,
\be \tilde P^{<(>)}(k_x,k_y) = \frac{1}{2} \left ( 
\frac{\tilde B_z^{\rm ph}(k_x,k_y)}{k_h^2} 
+(-) 
\frac{{\rm i} \kvec_h \cdot \tilde{\bvec}_h^{\rm ph}(k_x,k_y)}{k_h^3}
\right ) ~, \label{eqn:p_fft_pm} \ee
which can then be Fourier transformed to yield $P^<(x,y,z)$ and $P^>(x,y,z)$.
While we used continuous Fourier transforms here, the translation to 
discrete Fourier transforms is straightforward, with the notable caveat for finite domains that harmonic contributions to these potentials, consistent with assumptions made about BCs on the domain's perimeter, must be considered. 
%


\subsection{The Structure of Currents \& Boundary Conditions}
\label{subsec:} 

The formalism above employs the observed horizontal poloidal and normal fields at the photosphere as separate BCs, in Equations (\ref{eqn:ptd_dpdz_ltgt}) and (\ref{eqn:ptd_lapp_ltgt}).
Understanding how currents above and below the photosphere produce each of these field components in idealized cases can yield insights into relationships between observed structures in photospheric magnetic fields and the structures of their source currents.
As a starting point, we decompose the current density into horizontal and vertical components, $\jvec = \jvec_h + J_z \hatz$. 
%
While the poloidal-toroidal decomposition does not
split currents in this way, doing so can nonetheless aid understanding how currents affect the Neumann and Dirichlet BCs for the poloidal potentials.

\begin{figure}[ht]
\includegraphics[width=6.0in]{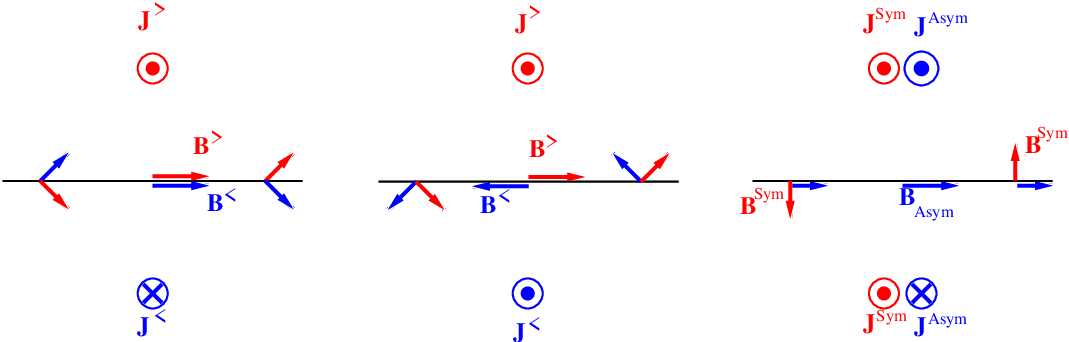}
\caption{\textsl{ Left: A current above the photosphere, $\jvec^>$
    (red vector pointing toward the reader) produces a magnetic field,
    $\bvec^>$, shown by the red vectors lying in the plane of the
    figure. (In this and other panels, all magnetic field vectors’
    colors match the color of their source currents.)  An equal but
    opposite current below the photosphere, $\jvec^<$ (blue vector
    pointing away from the reader), produces a magnetic field,
    $\bvec^<$, shown by blue vectors lying in the plane of the figure.
    These antisymmetric currents produce no normal field at the
    photosphere.  Adopting a coordinate system with $+x$ to the right
    in this figure, notice that these currents produce a nonzero
    $\partial_x B_x$ at the photosphere, implying that $(\nabla_h
    \cdot \bvec_h) \ne 0$.  Middle: Symmetric currents above and below
    the photosphere produce a normal magnetic field at the
    photosphere, but no horizontal field there.  Right: A coronal
    current without a matching sub-photospheric counterpart can be
    represented as the sum of symmetric and antisymmetric currents,
    with $|\jvec^{\rm Sym}| = |\jvec^{\rm Asym}|$; the $\jvec$ vectors
    shown are meant to be co-spatial. Such a configuration produces
    both normal and horizontal photospheric fields.}}
\label{fig:off_p-sphere}
\end{figure}

First, we consider the contribution of horizontal currents to 
to  $\bvec^{\rm ph}(x,y) = \bvec(x,y,0)$; we consider the contribution of vertical currents in the next paragraph.
Figure \ref{fig:off_p-sphere} illustrates components of the magnetic field produced at the photosphere due to  horizontal currents 
above and below it. 
The currents depicted should be understood to be short, horizontal portions of larger current 
systems that contain both vertical and horizontal segments.
The magnetic field vectors' colors match the color of their source currents.  
A horizontal current flowing in the corona will produce a magnetic field with both normal and horizontal components at the photosphere.  
As the figure's left panel illustrates, for the
special case of horizontal currents that are antisymmetric across the photosphere, no normal magnetic field is produced at the photosphere.
This property of mirror/image currents has been
exploited in coronal magnetic field modeling (e.g.,
\citealt{Sakurai1981a, Wheatland2007}).
%
(\citealt{Sezginer1990} extend the concepts of mirrored horizontal currents in Cartesian geometry to azimuthal current rings in spherical geometry. 
\citet{Titov2025} generalized the treatment of interior mirror currents for arbitrary exterior line and volume currents.)
The middle panel shows the situation with currents that are symmetric across the photosphere: in this case, a normal field is produced at the photosphere, but no horizontal field is produced there.
Thus, antisymmetric, off-photosphere horizontal currents produce the photosphere's horizontal poloidal field, and symmetric, off-photosphere, horizontal currents produce the photosphere's normal field.

The anti- and symmetric configurations considered above are special cases, but these considerations apply to more general field configurations.  
The right panel of Figure \ref{fig:off_p-sphere}, in which $|\jvec^{\rm Sym}| = |\jvec^{\rm Asym}|$ and the $\jvec$ vectors are meant to be co-spatial, shows a configuration with nonzero current in the corona, but zero current below it.
The logic used in the right-panel example can be applied for any horizontal current at any point, illustrating how an arbitrary current density can be represented as the sum of currents that are symmetric and antisymmetric about the $z=0$ plane. (In fact, any vector field can be decomposed this way.)
Thus, even in a generic case, with horizontal currents producing both horizontal and vertical magnetic fields at the photosphere,
we can ascribe (i) the horizontal variations of the horizontal photospheric field in Equation (\ref{eqn:ptd_dpdz_ltgt}) as due to antisymmetric currents, and (ii) the vertical photospheric field in Equation (\ref{eqn:ptd_lapp_ltgt}) as due to symmetric currents.

Second, we consider the contributions to $\bvec^{\rm ph}(x,y)$ from
$J_z(x,y,z)$ when $z \ne 0$.
The cross product in the numerator of Equation (\ref{eqn:bs_law})
implies that the vertical component of the current at any source
point does not produce any vertical magnetic field at the photosphere.
A vertical current segment does produce a horizontal
photospheric field, but, for the real photosphere, the horizontal photospheric field produced by a current system that does not penetrate the photosphere must have zero circulation on the surface, per the integral form of Amp\`ere's law 
($\mu_0 I_{\rm enc} = \oint \bvec \cdot d\Bell$).  
(Note, however, that a nonzero photospheric circulation can arise from off-photosphere vertical currents in field models in doubly-periodic Cartesian domains; see the discussion in \citealt{Schuck2026}.) 

%


%
%
%

\subsection{Gauss's Separation with Scalar Potentials}
\label{subsec:potential_separation}



In many solar magnetic modeling applications, current-free magnetic fields $\bvec^{\rm (P)}$, for which $\nablavec \times \bvec^{\rm (P)} = 0$, are represented as the gradient of a scalar potential, with $\bvec^{\rm (P)} = -\nablavec \psi$,
rather than in terms of the PTD potentials.  
Because magnetic fields are divergence-free, current-free fields also obey the 3D Laplace's equation, $\nablavec \cdot \bvec^{\rm (P)} = -\nabla^2 \psi = 0$, for which solution methods are well known (e.g., \citealt{Jackson1975}). 
Accordingly, in this section, we present the Gaussian separation in terms of $\bvec^<(x,y,0) = -\nablavec \psi^<(x,y,0)$ and $\bvec^>(x,y,0) = -\nablavec \psi^>(x,y,0)$. 
We note that, depending upon the numerical methods employed, a magnetic field derived from the gradient of a scalar-potential solution to Laplace's equation might not obey the divergence-free condition to machine precision. 
In terms of these potentials, Gauss's decomposition of the photospheric vector magnetic field in Equation (\ref{eqn:b_decomp}) can be written
\be \bvec^{\rm ph}(x,y) =
\nablavec_h \times [ T(x,y,0) \hatz ]
- \nablavec \psi^<(x,y,z) \vert_{z=0} - \nablavec \psi^>(x,y,z) \vert_{z=0} ~,
\label{eqn:b_decomp3} \ee
where, as above, currents flowing across $z=0$ generate the toroidal term, 
currents flowing below the photosphere produce $\bvec^<(x,y,z)$ equal to 
\be \bvec^<(x,y,0) = -\nablavec_h \psi^<(x,y,0) - \partial_z \psi^<(x,y,z)\vert_{z=0} \hatz
~, \quad z \in [0, \infty) \label{eqn:blt_psi} \ee
and currents flowing above the photosphere produce $\bvec^>(x,y,0)$ equal to 
\be \bvec^>(x,y,0) = -\nablavec_h \psi^>(x,y,0) - \partial_z \psi^>(x,y,z) \vert_{z=0} \hatz ~, \quad z \in (-\infty, 0]
~. \label{eqn:bgt_psi} \ee
Normal derivatives at $z = 0$ should be understood to be one-sided, taken from the direction in which the field is potential (i.e., from above for $\psi^<$ and from below for $\psi^>$), as is typical for solutions to Laplace's equation on a given domain \citep{Jackson1975}.
As above, taking the horizontal divergence of Equation (\ref{eqn:b_decomp3}) gives one equation (with no contribution from the $T$ term),
\be (\nablavec_h \cdot \bvec_h^{\rm ph}) = -\nabla_h^2 \psi^< - \nabla_h^2 \psi^> 
~. \label{eqn:dirichlet} \ee
%
Also as above, the linearity of the Biot-Savart law in its source currents implies 
%
\be B_z^{\rm ph}(x,y) = 
- \partial_z \psi^<(x,y,z) \vert_{z=0}  
- \partial_z \psi^>(x,y,z) \vert_{z=0}  \label{eqn:neumann} ~. \ee
The divergence-free condition implies $\nablavec \cdot \bvec^< = \nablavec \cdot \bvec^> = 0$, and the potentials $\psi^<$ and $\psi^>$ both
obey the 3D Laplace's equation, 
\bea  \nabla^2 \psi^< &=& 0 ~, z \in [0, \infty) ~, \label{eqn:laplace<} \\
        \nabla^2 \psi^> &=& 0 ~,  z \in (-\infty, 0] ~. \label{eqn:laplace>} \eea 
As above, normal derivatives in these expressions should again be understood as one-sided (from above for $\psi^<$ and from below for $\psi^>$), as each scalar field only obeys Laplace's equation within its specified domain. 
(We note that, in contrast to the differing domains of $\psi^<$ and $\psi^>$ here, the poloidal potentials $P^<$ and $P^>$ can be defined over the full range of $z$, per Equations~(\ref{eqn:PLT})\--(\ref{eqn:PGT}) in the Appendix.)
If we were solving for a single potential that solved Laplace's equation, 
$(\nablavec_h \cdot \bvec_h^{\rm ph})$ 
would supply a Dirichlet BC for a potential function $\psi^D$ \citep{Welsch2016}, 
and 
$B_z$
would form a Neumann BC for a potential function $\psi^N$.   
Either one of these boundary conditions is sufficient to completely specify a solution to Laplace's equation (in either the interior or exterior volume), and in general a single solution to Laplace's equation cannot satisfy arbitrary functions for both BCs (e.g., \citealt{Jackson1975}).  
In this case, however, we have two potential functions, $\psi^<$ and $\psi^>$, which, together, must satisfy both boundary conditions.

%
%

The potentials $\psi^<$ and $\psi^>$ can be expressed in terms of Fourier
transforms as
\be \psi^{<(>)}(x,y,z) = \frac{1}{(2\pi)^2}
\int_{-\infty}^{+\infty} {\rm d}k_x \int_{-\infty}^{+\infty} {\rm d}k_y \,
\tilde \psi^{<(>)}(\kvec) {\rm e}^{{\rm i}k_xx + {\rm i}k_yy -(+) k_h z}
~, \label{eqn:chi^pm_def} \ee
where the tilde denotes the Fourier transform of a function.
For each potential to solve Laplace's equation, we require $k_h \equiv
|\sqrt{k_x^2 + k_y^2}|$.   Because $\psi^< \to 0$ as $z \to +\infty$, each of its Fourier modes decays as exp$(-k_h z)$, while $\psi^> \to 0$ as $z \to -\infty$, implying that each of its modes decays as exp$(+k_h z)$.
In terms of these Fourier transforms,  Equation (\ref{eqn:neumann}) becomes  
\bea B_z^{\rm ph}(x,y) &=& 
-\partial_z \psi^< \vert_{z=0} - \partial_z \psi^>\vert_{z=0} \label{eqn:neum_source} \\
&=& \frac{1}{(2\pi)^2} \int_{-\infty}^{+\infty} {\rm
  d}k_x \int_{-\infty}^{+\infty} {\rm d}k_y \, [ \tilde \psi^<
(k_x,k_y) - \tilde \psi^> (k_x,k_y) ] \, k_h \, e^{{\rm i}k_xx + {\rm
    i}k_yy} ~. \label{eqn:neum_source_fft} \eea
Notice that the differing $z$-dependence of $\psi^<$ and $\psi^>$ implies that this condition involves a {\em difference} of their spectral functions.
Denoting the Fourier transform of $B_z^{\rm ph}(x,y)$ as $\tilde B_z^{\rm ph}(k_x,k_y)$, we have
\be \tilde \psi^< (k_x,k_y) - \tilde \psi^> (k_x,k_y)
= \frac{
\int_{-\infty}^{+\infty} {\rm d}x
\int_{-\infty}^{+\infty} {\rm d}y \, B_z^{\rm ph}(x,y)
    {\rm e}^{-{\rm i}k_xx - {\rm i}k_yy} }{k_h}
= \frac{\tilde B_z^{\rm ph}(k_x,k_y)}
{k_h} 
~. \label{eqn:neumann_psi} \ee
In terms of the Fourier transforms of $\psi^<$ and $\psi^>$, Equation (\ref{eqn:dirichlet}) becomes 
\bea
\nablavec_h \cdot \bvec_h^{\rm ph}(x,y) 
&=&
\nablavec_h \cdot \left ( \frac{1}{(2\pi)^2} \int_{-\infty}^{+\infty} {\rm d}k_x
\int_{-\infty}^{+\infty} {\rm d}k_y \, 
\tilde{\bvec}_h^{\rm ph}(k_x,k_y) {\rm e}^{ {\rm i}k_xx + {\rm i}k_yy} \right ) 
  \\
  %
  &=&
 \frac{1}{(2\pi)^2} \int_{-\infty}^{+\infty} {\rm d}k_x
\int_{-\infty}^{+\infty} {\rm d}k_y \, 
    {\rm i} \kvec_h \cdot \tilde{\bvec}_h^{\rm ph}(k_x,k_y)
    {\rm e}^{ {\rm i}k_xx + {\rm i}k_yy}
\eea
%
\be = 
\frac{1}{(2\pi)^2} \int_{-\infty}^{+\infty} {\rm d}k_x \int_{-\infty}^{+\infty}
{\rm d}k_y \, ( \tilde \psi^>(k_x,k_y) + \tilde \psi^<(k_x,k_y)) (k_x^2 + k_y^2) {\rm e}^{{\rm i}k_xx + {\rm i}k_yy}
~, \label{eqn:dirichlet_bc}
\ee
%
where $\tilde \bvec_h^{\rm ph}(k_x, k_y)$ is the 2D Fourier transform of $\bvec_h^{\rm ph}(x,y).$
The previous two equations imply that
\be \tilde \psi^<(k_x,k_y) + \tilde \psi^>(k_x,k_y) 
= \frac{{\rm i} \kvec_h \cdot \tilde{\bvec}_h^{\rm ph}(k_x,k_y)}{k_h^2}
~. \label{eqn:dirichlet_psi} \ee
%

%
%

While Equation
(\ref{eqn:neumann_psi}) involved a difference of $\tilde \psi^>$ and
$\tilde \psi^<$, Equation (\ref{eqn:dirichlet_psi}) involves their {\em sum}.  
%
Again we see this
key aspect of Gauss's separation method: the structure of
the horizontal and vertical components of the field in the separation
surface constrain the sum and the difference, respectively, of the
interior and exterior potentials' spectral functions.
%
%
%
The sum and the difference of Equations (\ref{eqn:neumann_psi}) and
(\ref{eqn:dirichlet_psi}) can then be used to find both the spectral
functions, $\tilde \psi^<$ and $\tilde \psi^>$,
\be \tilde \psi^{<(>)}(k_x,k_y) = \frac{1}{2} \left ( 
\frac{{\rm i} \kvec_h \cdot \tilde{\bvec}_h^{\rm ph}(k_x,k_y)}{k_h^2}
+(-) \frac{\tilde B_z^{\rm ph}(k_x,k_y)}{k_h} \right ) ~, \label{eqn:chi_fft_pm} \ee
which can then be Fourier transformed to yield $\psi^<(x,y,z)$ and $\psi^>(x,y,z)$.

We now show that non-periodic solution methods can be employed to find $\psi^<$ and $\psi^>$.
Below Equation (\ref{eqn:laplace>}), we briefly discussed the two additional potential functions, $\psi^D$ and $\psi^N$, which separately satisfy Laplace's equation for Dirichlet and Neumann BCs, respectively.  
We can write the Fourier solutions for these potentials, arbitrarily choosing the exterior solution for both:
\be \psi^{D(N)}(x,y,z) = \frac{1}{(2\pi)^2}
\int_{-\infty}^{+\infty} {\rm d}k_x \int_{-\infty}^{+\infty} {\rm d}k_y \,
\tilde \psi^{D(N)}(\kvec) {\rm e}^{{\rm i}k_xx + {\rm i}k_yy -k_h z}
~. \label{eqn:chi^DN_def} \ee
From the nature of Dirichlet and Neumann boundary conditions, it must be true that 
\bea 
\tilde \psi^N (k_x,k_y) &=& \tilde \psi^<(k_x,k_y) - \tilde \psi^>(k_x,k_y)  \\
\tilde \psi^D (k_x,k_y) &=& \,\, \tilde \psi^<(k_x,k_y) + \tilde \psi^>(k_x,k_y) 
~, \eea
meaning  
\be
\tilde \psi^{<(>)}(k_x,k_y) = \frac{1}{2} \left ( 
\tilde \psi^{D}(k_x,k_y) +(-) \, \tilde \psi^{N}(k_x,k_y) \right )
~. \label{eqn:non_periodic_FFT} \ee
The linearity of the Fourier transform then implies that this relation between the {\em spectral functions} also applies to the {\em potentials themselves} at $z = 0$,
\be
\psi^{<(>)}(x,y,0) = \frac{1}{2} \left ( 
\psi^{D}(x,y,0) +(-) \, \psi^{N}(x,y,0) \right )
~. \label{eqn:non_periodic} \ee
For the continuous Fourier approach used here, the minimum wavenumber $k_{\rm min} = 0$ corresponds to a domain size $L = \infty$. Hence, for localized photospheric magnetic fields, 
any solution method for $\psi^D$ and $\psi^N$ yielding fields that vanish at infinity can be employed, and the assumption of periodicity in the horizontal directions is not necessary.
\citet{Welsch2016} presented finite-element methods to 
derive both potential functions appearing in Equation (\ref{eqn:non_periodic}) for localized sources in non-periodic configurations.  Spectral methods can be also adapted for non-periodic boundaries; there are codes for such cases in the CICCI library.

As with the separation in terms of PTD potentials in Section \ref{subsec:ptd_separation} above, however, we translate the continuous Fourier transforms used here to discrete Fourier transforms in what follows.  Again, the reader is urged to
bear in mind that on finite domains one must consider the contribution of harmonic terms to these potentials, discussed near Equation \ref{eqn:bh_match}. 
%
Knowledge of $\psi^<(x,y,z)$ and $\psi^>(x,y,z)$ can then be used to determine
$\bvec^<(x,y,z)$ for $z \ge 0$ and $\bvec^>(x,y,z)$ for $z \le 0$,
including the part of the photospheric field due to interior currents,
$\bvec^<(x,y,0)$, and the part due to coronal currents,
$\bvec^>(x,y,0)$.

We remark that \citet{Welsch2016} found that the two potential fields
determined separately from Neumann and Dirichlet boundary conditions 
($\bvec^N(x,y,0) = -\nablavec \psi^N(x,y,0)$ and 
$\bvec^D(x,y,0) = -\nablavec \psi^D(x,y,0)$ respectively) 
taken from active-region vector magnetograms were inconsistent with
each other.  
This inconsistency only arises from assumption that the coronal field is truly potential; without that assumption, the inconsistency ceases to exist.

%
%


%

%

Inspection of $P^<$ and $P^>$ in Equations (\ref{eqn:blt_ptd})  and (\ref{eqn:bgt_ptd}), respectively, and $\psi^<$ and $\psi^>$ in Equations (\ref{eqn:blt_psi}) and (\ref{eqn:bgt_psi}), respectively, make clear that the poloidal and scalar potentials are related, {\em where each field is potential}, via  
\bea 
\partial_z P^< &=& -\psi^< ~, \quad z > 0 \\
\partial_z P^> &=& -\psi^> ~, \quad z < 0 ~.
\eea 
These relationships {\em only hold} where the field is potential --- in current-carrying regions, vertical derivatives of the poloidal potential contribute to currents.

For completeness, we note that the spectral function $\tilde T(k_x,k_y)$ for the toroidal potential $T(x,y,0)$ is given from the Fourier transform of Equation (\ref{eqn:jzi}), 
\be \tilde T(k_x,k_y) = +\mu_0 \tilde J_z(k_x, k_y)/(k_x^2 + k_y^2) ~, \label{eqn:tilde_t} \ee
where $\tilde J_z(k_x, k_y)$ is the 2D Fourier transform of $J_z(x,y,0).$
For the Fourier methods presented here, routines in 
IDL \citep{welsch_2026_zenodo_IDL}
and 
Python 3 \citep{Welsch2026_github}
using discrete Fourier transforms have been made available online,
along with brief documentation, a sample input data file, and example results.
%




\section{Gauss's Decomposition of AR 10930}
\label{sec:10930}

To illustrate results of using the decomposition in Equation (\ref{eqn:b_decomp}) 
with data, we apply it to the observed photospheric magnetic field of AR 10930.
This active region was the focus of a detailed comparison of several
NLFFF coronal field models by \citet{Schrijver2008}, so several models
of its coronal currents have been made.  It possessed a strongly
sheared central polarity inversion line 
that was the site of initial brightenings in 
%
an X3.4 flare that occurred on
2006/12/13, with a peak in GOES soft X-rays at 2:40 UT.
The region was relatively near disk center at this time, with the
USAF/NOAA Solar Region Summary issued at 24:00 UT on 2006/12/12
reporting its position as S06W21.
The magnetogram we have used was prepared for that study,
and has been posted on Zenodo \citep{Schrijver2026}.
%
%
The vector magnetic field was derived from observations by the
SpectroPolarimeter (SP; \citealt{Lites2013}) instrument on the Solar
Optical Telescope (SOT; \citealt{Tsuneta2008, Suematsu2008,
  Ichimoto2008, Shimizu2008}) aboard the Hinode satellite
\citep{Kosugi2007}, 
%
using rasters over about 45 minutes starting at
20:30 UT on 2006/12/12.

To create the magnetogram used by \citet{Schrijver2008}, the SP field
was embedded within a larger magnetogram derived from observations of
the line-of-sight field made by the Michelson-Doppler Imager
\citep{Scherrer1995}.
(The boundary between the SP magnetogram and the
larger-field-of-view MDI magnetogram exhibits some jumps in field strength. Discontinuities like this will produce high-wavenumber artifacts, which could be avoided by using full-disk magnetograms.  The fields in the AR's core that are our focus are, however, well away from this boundary.)
%
%
The data were interpolated onto a uniform grid with pixels
0.63$\arcsec$ on a side, corresponding to 460 km at the solar
photosphere.
All potentials derived here for AR 10930 were computed on (320
$\times$ 320) grids, which were then cropped around the central part of the
active region to focus on magnetic structure there.  

\begin{figure}[hp]
\includegraphics[angle=90,scale=0.8]{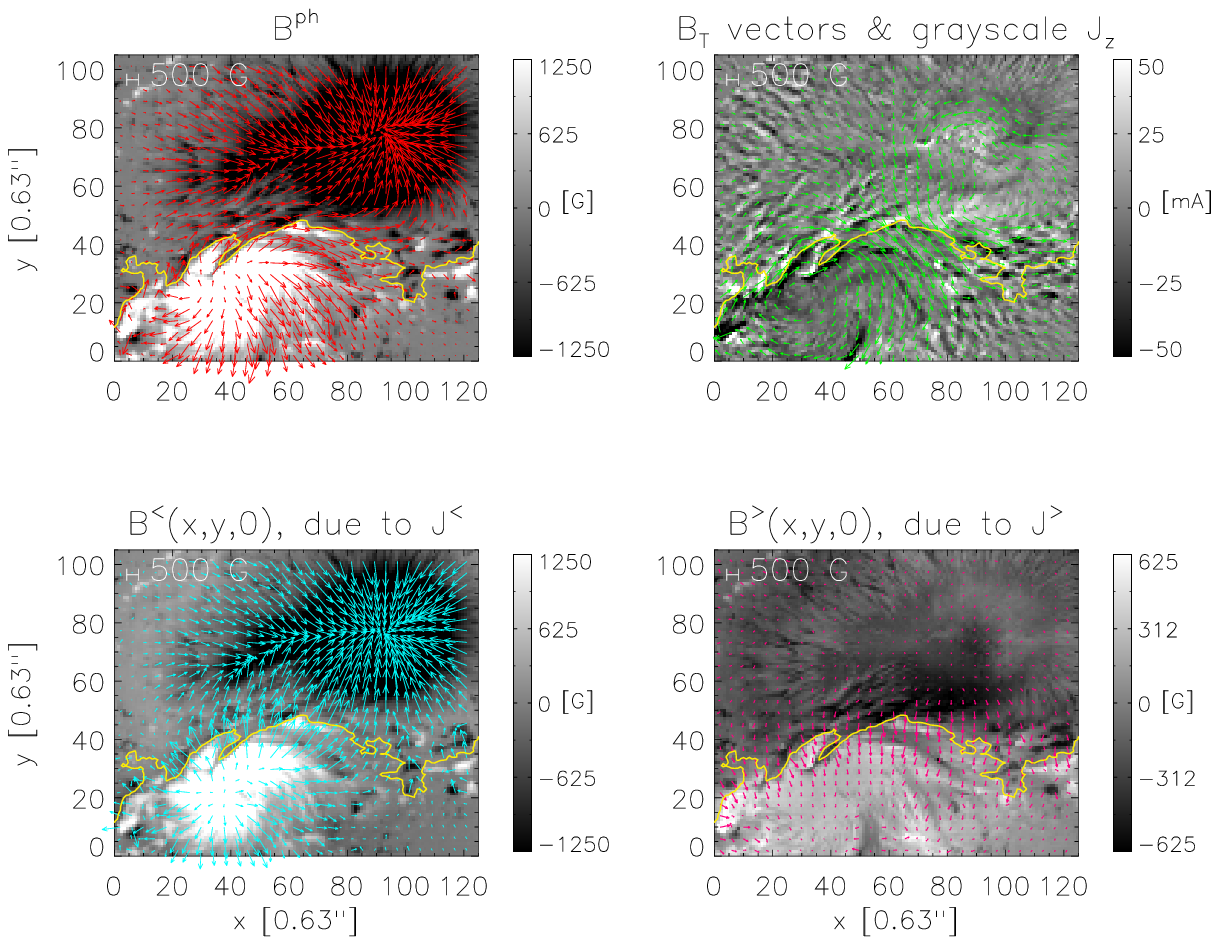}
\caption{\textsl{Top left: The magnetic field in the central region of AR 10930, observed by Hinode/SOT/SP, with rasters starting at 20:30 UT on 2006/12/12.  
Background grayscale shows $B_z$.
Yellow $B_z = 0$ contour in each panel shows the main PIL. Top right: Vectors show the horizontal field's toroidal component, and grayscale shows $J_z$ in mA. 
Bottom-left: Vectors show $\bvec^<(x,y,0)$, with $B_z^<(x,y,0)$ in background grayscale 
and $\bvec_h^<(x,y,0)$ as aqua vectors.  
Bottom right: $\bvec^>(x,y,0)$, with $B_z^>(x,y,0)$ in background grayscale
and $\bvec_h^<(x,y,0)$ as hot pink vectors. 
Only every fourth vector is plotted, to reduce clutter. 
}}
\label{fig:ar10930_blt_bgt_vects}
\end{figure}

In the top-left panel of Figure \ref{fig:ar10930_blt_bgt_vects}, we show 
the observed photospheric magnetic fields in the central portion of AR 10930.  
%
%
Green vectors in the top-right panel show the toroidal component of the horizontal field, $\bvec_T$, with $J_z$ in background grayscale (saturated at 50 mA/m$^2$).
Aqua vectors in the bottom-left panel show
$\bvec_h^<(x,y,0)$, with $B_z^<(x,y,0)$ in background grayscale
(saturated at $\pm$ 1250 G).
Pink vectors in the bottom-right panel show 
$\bvec_h^>(x,y,0)$, with $B_z^>(x,y,0)$ in
background grayscale (saturated at $\pm$ 625 G).
For the background grayscale in this figure and all others in this
paper, black areas correspond to negative quantities, and white areas
correspond to positive.
In this figure, only every fourth vector is plotted, to reduce clutter.

In the top-left panel, counter-clockwise vorticity at the center of the strong, negative field in the upper-right portion of the frame (a sunspot) indicates a large-scale current flows upward there ($J_z > 0$), and clockwise vorticity in the strong, positive flux region toward lower left of the frame (also a sunspot) implies a large-scale current flows downward ($J_z < 0$) there.  The region's main PIL is overplotted as a yellow, $B_z = 0$ contour in this panel and others in this figure. 
The top-right panel shows that strong magnetic shear along the main PIL contour is due to the toroidal field, $\bvec_T$, and is thus caused by the large-scale vertical current systems described above. Smaller-scale, more intense vertical currents are also present along the PIL, but these do not appear to influence the global structure of $\bvec_T$.

There is some debate about how the strongly sheared PIL developed. It could have arisen from vertical currents flowing within a single, large-scale flux system that emerged to form this region.
Alternatively, this active region might have developed from the collision of the newly emerging flux system with remnants of an older region (suggested by G. Chintzoglou in a private communication in 2024), in which case the vertical currents that produce the sheared field originated in separate flux systems. 

In the bottom row, the factor-of-two difference in saturation levels between left and
right panels shows that, on average, vertical fields arising from
coronal currents are significantly weaker than vertical fields from
interior currents.
In the bottom-right panel, an overall dipolar structure in $B_z^>(x,y,0)$ can
be seen, with coronal currents producing negative flux north of the
large-scale PIL and positive flux south of it.

\subsection{$\bvec^<$ and $\bvec^>$ Suggest Parallel $\jvec_h^<$ and $\jvec_h^>$}
\label{subsec:jhgt_jhlt}

Near the PIL in the bottom-row panels of Figure \ref{fig:ar10930_blt_bgt_vects}, the signs of $B_z^<$ and $B_z^>$ match in corresponding areas.
Based on the middle panel of Figure \ref{fig:off_p-sphere}, the similarity in $B_z^<$ and $B_z^>$ suggests that substantial parts of the $\jvec_h^<$ and $\jvec_h^>$ that produce these fields are symmetric across $z=0$, with both flowing roughly leftward (eastward on the Sun). 

In the bottom-left panel of Figure \ref{fig:ar10930_blt_bgt_vects}, it can be seen that $B_y^<$ generally points away from the
positive sunspot, and in many areas toward the negative sunspot. Thus, across the PIL between the sunspots, the horizontal field exhibits the ``normal'' orientation for a generic potential field.  In contrast, $B_y^>$ in the bottom-right panel points away from the negative spot across the main PIL, toward the positive spot
--- i.e., an ``inverse'' orientation \citep{Low1995, Kuperus1974}
with respect to a generic potential field, or a ``bald-patch'' configuration  \citep{Titov1993}.  
Very similar near-PIL structure in $\bvec_h^>$ was observed by \citet{Schuck2022} in AR 12673. 
Again, based on the middle panel of Figure \ref{fig:off_p-sphere}, the opposite directions of $B_y^<$ and $B_y^>$ across the PIL suggest that substantial parts of the $\jvec_h^<$ and $\jvec_h^>$ that produce these fields are symmetric across $z=0$, again flowing roughly leftward. 

Because parallel currents attract, it is plausible that symmetric-current configurations like that observed here are more stable than antisymmetric-current systems, and are thus more likely to be observed.
We expect that, if anti-aligned current systems were observed, they would tend to be evolving rapidly --- perhaps toward a more stable, nearly parallel-current configuration, or perhaps toward more disruptive evolution.  

\subsection{Statistically, $\bvec^<$ is stronger than $\bvec^>$}
\label{subsec:bgt_blt_stats}

From the distribution of field strengths in an area of plage in this
magnetogram southeast of the positive spot, \citet{Welsch2015}
estimated the noise level in the components of $\bvec$ to be circa 35
G.  Among pixels with observed vertical field $|B_z| > 50$ G, the
median values of $|B_z^<|$ and $|B_z^>|$ are both about 150 G, but the
means of $|B_z^<|$ and $|B_z^>|$ are 360 G and 180 G, respectively.
Among this same subset of pixels, the total unsigned flux in $B_z^<$ 
is $2.7 \times 10^{22}$ Mx, while in $B_z^>$ the total unsigned flux is
$1.4 \times 10^{22}$ Mx --- about half as large.
Because magnetic energy, an important physical quantity, scales quadratically in field strength, comparisons involving $(\bvec^<)^2$ and $(\bvec^>)^2$ are also interesting.  
Summing these squares over the magnetogram, we find $\sum (\bvec^<)^2 = 3.1 \times 10^{10}$ G$^2$ and $\sum (\bvec^>)^2 = 3.1 \times 10^{9}$ G$^2$ --- a difference of a factor of ten, implying $|\bvec^<|$ tends to be more than a factor of two larger than $|\bvec^>|$ in the strong-field regions that contribute more heavily in these sums of squares.


\section{Gauss's Decomposition of AR 11158}
\label{sec:11158}

We now use Gauss's method to decompose photospheric fields in another active region, AR 11158.  
Like AR 10930, this region developed a strongly sheared central PIL and
produced an X-class flare.  Notably, this AR
appeared to form from the nearby emergence of two separate,
large-scale flux tubes \citep{Chintzoglou2013}.
\citet{Chintzoglou2019} analyzed the evolution of shear along the
central PIL as nearby footpoints from the two different tubes appeared to
collide at the photosphere, which they dubbed collisional shearing.

Here, we analyze magnetic structure in AR 11158 at 01:36 UT on 2011/02/15, 
just before the X2.2 flare that started at 01:44 UT in GOES soft X-ray observations. 
%
%
We used data available online via the SDO Joint Science Operations Center at Stanford University (JSOC; http://jsoc.stanford.edu/) in the ``cgem.pdfi\_input'' series, prepared for the Coronal Global Evolutionary Model (CGEM) project \citep{Hoeksema2020}.   
This series contains vector magnetic field, Doppler velocity, optical-flow velocity, and electric field estimates for many active regions.  
Notably, the fields of view in CGEM data series for many ARs encompass larger areas than HARP or SHARP series, so CGEM magnetograms are often surrounded by larger areas lacking AR magnetic fields compared to HARP/SHARP magnetograms.
Having a large ``moat'' free of AR fields surrounding our AR of interest is valuable, because this reduces the significance of harmonic terms in the Gaussian separation within the AR.
The input data were (600 $\times$ 600) arrays, and all intermediate calculations used to create the images that we show were computed on the full FOV and then cropped to focus on areas of interest.  

The data in the CGEM series are interpolated onto a {\em plate-car\'ee} (``square-flat'' in French) or equirectangular grid, i.e., pixels have constant angular widths in longitude and latitude.  Both widths were chosen to be $0.03^\circ$ in this series, to match HMI's disk-center pixel size, corresponding to about 360 km.   
It should be noted that 
\emph{plate-car\'ee} projection is neither authalic (area-preserving; e.g., \citealt{DeForest2007}) nor conformal (shape-preserving), so it both distorts areas and directions.  These distortions will bias derivative operators (e.g., $\nablavec_h \cdot$ and $\nablavec_h \times$).  To mitigate such distortions, the JSOC reprojection for a given FOV is local, with the origin of the {\em projection} coinciding with that FOV's center. 
(In contrast to the origin of the projection, the coordinate system's origin is Stoneyhurst \citep{Thompson2006}.)

%

\begin{figure}
\includegraphics[width=7.0in]{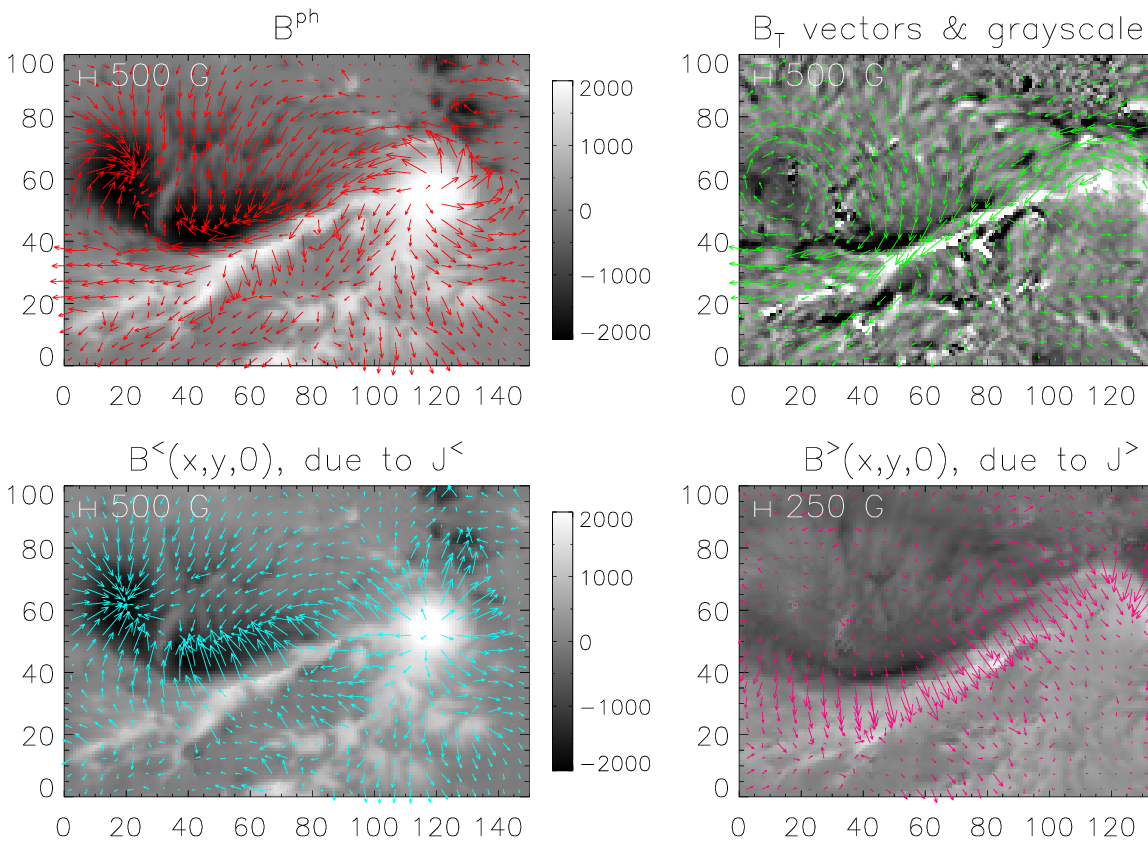}
\caption{\textsl{ Top left: Grayscale shows $B_z$ in 
    the central portion of AR 11158 at 01:36UT on 2011/02/15, shortly before the onset of the X2.2 flare at 01:44 UT. 
    Red vectors show the observed horizontal magnetic field.
    Top right: Green vectors show the toroidal photospheric magnetic field, $\bvec_T$, with $J_z$ [mA m$^{-2}$] shown in grayscale.  
    A large-scale upward (downward) current is present in vortical fields at frame right (left), which coincides with an area of positive (negative) flux. Smaller-scale, more intense currents are present along the sheared PIL.  Note that $\bvec_T$ makes the dominant contribution to the shear component of the field at the PIL. Bottom left: Vectors of $\bvec_h^<(x,y,0)$, with $B_z^<(x,y,0)$ in grayscale.  
    Bottom right: Vectors of $\bvec_h^>(x,y,0)$, with $B_z^>(x,y,0)$ in grayscale.  
    (Note smaller vector scale in this panel.)  
    Both the large-scale, coherent patterns in $\bvec^<(x,y,0)$ and $\bvec^>(x,y,0)$ near the PIL are consistent with large-scale currents, in the interior and in the corona, respectively, flowing along the PIL, mostly leftward and tilted toward the bottom of the frame.
    For all vector fields shown, only every fifth vector is plotted, to reduce clutter.}}
\label{fig:ar11158_blt_bgt}
\end{figure}

In Figure \ref{fig:ar11158_blt_bgt}, we show a zoomed-in view of the center of the AR, to clearly display the magnetic field's structure there.  
Red vectors in the top-left panel show the observed horizontal magnetic field in this sub-window, with $B_z$ in grayscale.   
Only every fifth vector is plotted, to reduce clutter.

Green vectors in the top-right panel show the pre-flare field's toroidal component, $\bvec_T(x,y,0)$, with the vertical current, 
$J_z(x,y)$, 
shown in grayscale. 
As with AR 10930, we can clearly see that vorticity in AR 11158 is present in $\bvec_T(x,y,0)$, and is associated with the vertical currents in the strong-field regions at either end of this sheared PIL (which are sunspots in continuum images).
As discussed by \citet{Chintzoglou2019}, the {\em intensification} of the sheared field along this PIL is likely due to the collision between its two large-scale flux systems.  We note, however, that the {\em origin} of these sheared fields must be vertical currents, $J_z$, flowing within each system. 
This point deserves emphasis, because sheared fields are a common feature of flare- and CME-productive regions: sheared fields are toroidal, not poloidal, meaning that {\em vertical photospheric currents are the physical cause of sheared fields along PILs.}  
Without modeling coronal fields (e.g., \citealt{Sun2012}), we cannot make concrete statements about what fraction of the vertical currents that cross the photosphere in these strong-field regions connect across this sheared PIL versus connecting to 
more distant parts of the active region.

The bottom-left panel shows 
$\bvec^<(x,y,0)$, and the bottom-right panel shows 
$\bvec^>(x,y,0)$.
Both the large-scale, coherent patterns in $\bvec_h^<(x,y,0)$ along the PIL and $B_z^<(x,y,0)$ straddling the PIL are consistent with a large-scale current flowing in the interior along the PIL, mostly leftward and tilted toward the bottom of the frame.  The structures of $\bvec_h^>(x,y,0)$ and $B_z^>(x,y,0)$ near the PIL are also consistent with a similarly large-scale current flowing along the PIL in the same direction, but above the PIL. 

%


The structure of fields observed in both this region and AR 10930 along their central PILs exhibits common features with near-PIL fields in AR 12673, described by \citealt{Schuck2022}.
First, the sheared field along PILs in all three regions is due to $\Bvec_T$, which arises from $J_z$ across the photosphere.
Second, the coherence and direction of $\bvec_h^>(x,y,0)$ along PILs in all three regions suggest that these field components arise from a relatively large-scale, horizontal current, $\jvec_h^>$, flowing above the photosphere, along the PIL. 
Third, from the directions of $\bvec_h^<(x,y,0)$ and $\bvec_h^>(x,y,0)$ along the PILs analyzed, the currents $\jvec_h^<$ and $\jvec_h^>$ near these PILs must flow roughly parallel to each other.

\subsection{Near-PIL Fields: Cross-PIL Gradients in $B_z$}
\label{subsec:pil_gradients}

Observers have reported significant statistical associations between strong gradients in the vertical fields across PILs --- or, equivalently, opposite-polarity fields in close proximity --- and the occurrence of flares \citep{Schrijver2007, Mason2010} and CMEs \citep{Falconer2003}. 
We note that these studies were performed prior to the ready availability of vector magnetograms with the advent of HMI, so line-of-sight (LOS) magnetic fields near disk center were often used as proxies for vertical fields. 
These strong-field PILs have been interpreted as signatures of horizontal currents: \citet{Schrijver2007} asserted that ``high-gradient, strong-field polarity-reversal lines in LOS magnetograms are in fact proxies of (near-)photospheric electrical currents.''

\begin{figure}
\vspace{-0.75in}
\includegraphics[width=8.0in,angle=90]{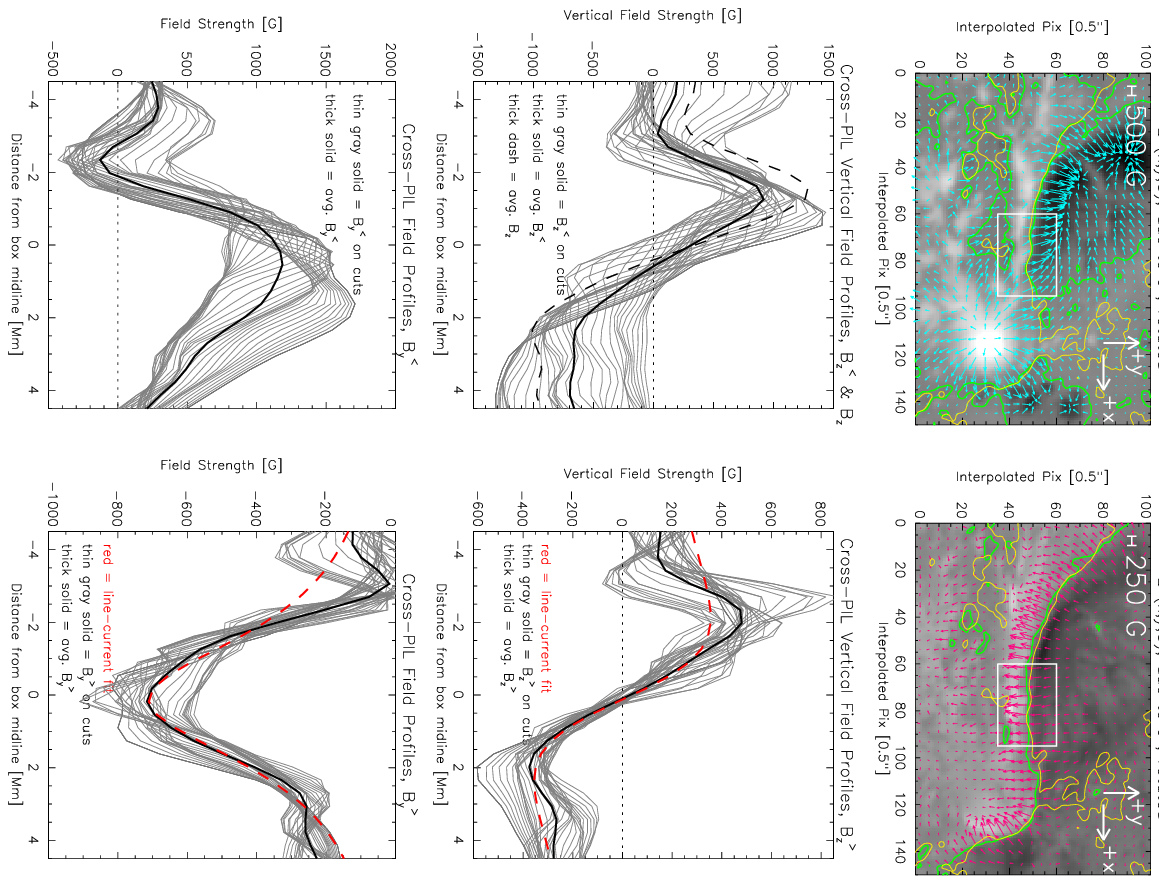}
\caption{\textsl{ Top left: A rotated view of a portion of the central PIL in AR 11158, showing $\bvec_h^<$ with aqua vectors, and $B_z^<$ in background grayscale. 
(saturated at $\pm 2000$ G). 
Yellow contour shows $B_z = 0$ contour, green contour shows $B_z^< = 0$ contour. White box outlines a near-PIL region, 36 pixels $\times$ 26 pixels, analyzed in panels below.  In this and the next panel, we define $+x$ toward page right and $+y$ toward the top of the page.
Top right: Analogous image for $\bvec^>$, showing $\bvec_h^>$ with pink vectors, $B_z^>$ in background grayscale (saturated at $\pm 1000$ G), and PILs of $B_z$ and $B_z^>$ in yellow and green, respectively. In both top-row panels, only every fourth vector is plotted, to reduce clutter.
    Middle left: gray curves show profiles of $B_z^<$ along vertical cuts across the white box in top-left panel, and thick black curve shows horizontal average of $B_z^<$ profiles.  Dashed black curve shows horizontal average of $B_z$ across the box.  Due to $B_z^>$, the gradient of $B_z$ across the PIL is steeper than the gradient of $B_z^<$ there.
    Middle right: gray curves show profiles of $B_z^>$ along vertical cuts across the white box in top-right panel, and thick black curve shows horizontal average of $B_z^>$ profiles. Dashed red-curve shows profile of $B_z^>$ from an $I_{\rm wire}$ = 800 GA line current at $z_{\rm s}$ = 2.25 Mm above the magnetogram surface.  
    Bottom left: gray curves show profiles of $B_y^<$ along vertical cuts across the white box in top-left panel, and thick black curve shows horizontal average of $B_y^<$ profiles.  
    Bottom right: gray curves show profiles of $B_y^>$ along vertical cuts across the white box in top-right panel, and thick black curve shows horizontal average of $B_y^>$ profiles. Dashed red-curve shows profile of $B_y^>$ from the same line-current model. Note that scales in the left and right columns of the middle and bottom rows differ.}}
\label{fig:ar11158_near_pil}
\end{figure}

In Figure \ref{fig:ar11158_near_pil}, we explore spatial variations of $\bvec^<, \,\, \bvec^>$, and $B_z$ across a portion of AR 11158's central PIL. (The differing vertical-axis scales should be noted.) Plots of $B_z^<, B_z^>$, and $B_z$ show that $B_z^>$, produced by horizontal current $\Jvec_h^>$ flowing above and along the PIL, increases $|\nablavec B_z|$ across the PIL.  This accords with the basic picture proposed by \citet{Schrijver2007}, and provides a physical underpinning for empirical associations between strong, cross-PIL gradients in $B_z$ and release of magnetic energy in flares / CMEs that is stored in currents flowing above the photosphere.


\subsection{Near-PIL Fields: A Line-Current Model for $\Jvec_h^>$}
\label{subsec:one_wire}

A line current can provide a simple model of the near-PIL horizontal current $\jvec_h^>$ that produces $\bvec_h^>$ there.  We apply this model to fields in the box shown in 
the upper-right panel of Figure \ref{fig:ar11158_near_pil}.
With the $x$ axis chosen to run horizontally along the box's midpoint, roughly paralleling the PIL, a current $I_{\rm wire}$ flowing to the left at source height $z_{\rm s}$ above the PIL along the $y = y_{\rm s}$ line produces field components
\bea 
B_{z,\rm{wire}}(y) &=& -\frac{\mu_0 I_{\rm wire} \, (y - y_{\rm s})}
{[(y - y_{\rm s})^2 + z_{\rm s}^2 ]} \label{eqn:bzwire} \\
B_{y,\rm{wire}}(y) &=& -\frac{\mu_0 I_{\rm wire}  \, z_{\rm s}}
{[(y - y_{\rm s})^2 + z_{\rm s}^2 ]}
\label{eqn:bywire} ~. 
\eea

Maxima in $|B_{z,\rm{wire}}(y)|$ occur at $(y - y_{\rm s}) = \pm z_{\rm s}$, so $z_{\rm s}$ can be estimated from the profile of averaged $B_z^>$ (thick, black solid line) in the middle-right panel to be about 2.25 Mm.  The location of the minimum in the profile of averaged $B_y^>$ in the bottom-right panel sets the distance between $y = 0$ (at the mindpoint of the box) and $y_{\rm s}$, about 0.13 Mm here.  
The depth of the average profile of $B_y^>$ can be used to estimate $I_{\rm wire}$, and the red, dashed curves shown in the middle- and bottom-right panels correspond to $I_{\rm wire}$ = 800 GA, where 1 GA = $10^9$ amp.  

The average profile of $B_z^<$ (thick, black solid line in the middle-left panel) only exhibits an extremum on one side of the PIL (about 2 Mm from it), so the wire-current model is less useful for modeling $\jvec_h^<.$ 
It is plausible that, for magnetic fields in the interior, convective forcing prevents active-region fields from forming coherent structures on length scales as long as those of chromospheric and coronal structures.

The modeled coronal current used here has several shortcomings.  The current distribution is singular, and thus unphysical.  The actual $\jvec_h^>$ certainly varies in ($x, y, z$), with currents flowing over a range of heights.
For example, the residual between $B_{z,\rm{wire}}(y)$ and the observed $B_z^>$ profiles might be ascribed to additional currents flowing at lower heights, which would contribute significantly to $B_z^>$ but only weakly to $B_y^>$, due to the source current height, $z_{\rm s}$, appearing in the numerator of equation (\ref{eqn:bywire}).
Despite the shortcomings of this simple model, the parameters given here provide some insight into the magnitudes and locations of currents that produce $\bvec^>$ near the PIL.  We remark that the large-scale, unipolar areas of significant $B_z^>$ farther from the PIL suggest that substantial additional currents flow at greater heights, but the structure of these currents is harder to directly infer from maps of $\bvec^>$.

\section{Conclusions: Discussion \& Future Work}
\label{sec:conclusions}

\subsection{Key Findings \& Implications}

In the two flare-productive active regions that we analyze here, we find that the  structure of fields near their central PILs exhibits features common to both regions and to near-PIL fields in AR 12673, analyzed by \citealt{Schuck2022}.
First, poloidal-toroidal decomposition of the horizontal photospheric field shows that the sheared field along the PILs in all three regions is due to $\Bvec_T$, which arises from $J_z$ across the photosphere.
On theoretical grounds, \citet{McClymont1989} noted that near-surface flows acting on already emerged fields are not expected to generate large currents in active regions. Subsequent observations by \citet{Leka1996} found that vertical photospheric currents arise from the emergence of current-carrying fields from the solar interior.  
Thus, the observed magnetic shear along PILs does not require either 
horizontal shearing motions (e.g., \citealt{Antiochos1999a}) or collisions between flux systems \citep{Chintzoglou2019}.   (Collisions between regions might, however, intensify magnetic shear along PILs.)

Second, the coherence and direction of $\bvec_h^>(x,y,0)$ along the analyzed PILs in all three regions suggest that this component of the field arises from relatively large-scale, horizontal currents, $\jvec_h^>$, flowing above the photosphere, along these PILs. 
Simplistically modeling $\bvec^>(x,y,0)$ along a portion of the central PIL in AR 11158 as arising from a line current, we found an $\sim$800 GA current flowing $\sim$2 Mm above the photosphere could reproduce aspects of the averaged $\bvec^>(x,y,0)$.

Third, in all three regions along the PILs analyzed, the reversed directions of $\bvec_h^<(x,y,0)$ and $\bvec_h^>(x,y,0)$ and same-sign distributions of $B_z^<(x,y,0)$ and $B_z^>(x,y,0)$ suggest that the near-PIL horizontal currents producing these fields, $\jvec_h^<$ and $\jvec_h^>$, must flow roughly parallel to each other.
Because parallel currents attract, parallel-current configurations
are likely more stable than misaligned-current configurations.  This suggests that the former are more likely to be observed than the latter, which should be prone to evolve toward a more stable configuration.
These observations of parallel currents above and below the photosphere also suggest that 
Lorentz forces in 
models of CME initiation that employ mirror (antisymmetric) currents in the interior to keep the normal photospheric magnetic field fixed, whether explicitly (e.g., \citealt{Forbes1991}) or implicitly (e.g., \citealt{Antiochos1999a}) are inconsistent with observations.

\subsection{The Importance of $\nablavec_h \cdot \bvec_h$}

\begin{figure}[ht]
\includegraphics[width=3.5in]{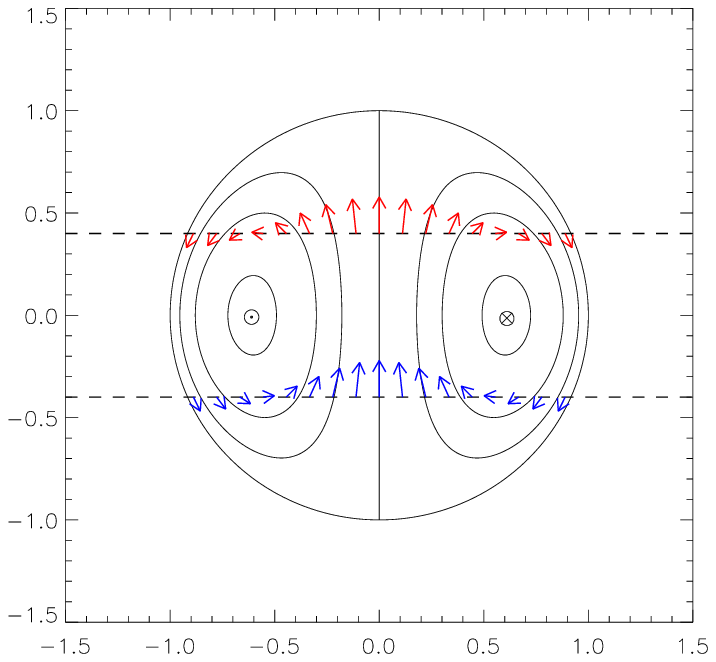}
\caption{\textsl{ Solid contours show poloidal magnetic field lines from a 
spheromak field.  The azimuthal field points out of the left half of the image, denoted by $\odot$ there, and into the right half of the page, denoted by $\otimes$ there. The dashed lines show two cross sections at $\pm z_0$, and
vectors show the poloidal field along each cross section.  Because $\jvec$ is parallel to $\bvec$ for this constant-$\alpha$, force-free field, these contours and vectors of $\bvec$ also show the corresponding quantities for $\jvec$.  
  Each cross section can be envisioned as a photospheric
  boundary condition, with an overlying coronal field.  The field
  above the upper cut, however, differs radically from the field above
  the lower cut, despite their matching distributions of $B_z$ and
  $J_z$: the latter contains a ``floating'' flux rope. The key
  photospheric imprint of this coronal flux rope in the lower cut is
  the ``inverse'' orientation of the horizontal field on its PIL.  The horizontal divergence, $\nablavec_h \cdot \bvec_h$, in the lower-cut field
  is opposite to that in the upper-cut field. Consequently, $\bvec^>$
  differs on the upper and lower cuts.}
\label{fig:spheromak}}
\end{figure}

While much research has focused on the structure of $B_z$ and $J_z$ (the latter from the horizontal curl of the horizontal field)
in flare- and CME-productive regions (e.g., \citealt{Janvier2014}),
Equation (\ref{eqn:ptd_dpdz_ltgt}) (or  (\ref{eqn:dirichlet}) in terms of $\psi$) show that $\nablavec_h \cdot \bvec_h$ contributes to $\bvec_h^>(x,y,0)$, so reflects aspects of coronal currents.   This
quantity can differ substantially between fields with similar
distributions of $B_z$ and $J_z$, with the difference being related to
currents in the corona.  To illustrate this, consider the vector field
in two cross sections of a constant-$\alpha$, force-free spheromak field (e.g., \citealt{Rosenbluth1979, Priest2000}), as illustrated in Figure
\ref{fig:spheromak}.  Because $\jvec$ and $\bvec$ are proportional to each other, field lines of $\bvec$ are also field lines {\bf of} $\jvec$.  Two cuts across the spheromak field, at heights $\pm z_0$, may be
envisioned as a photospheric boundary condition, with an overlying
coronal field.  
Despite these cuts having matching distributions of $B_z$ and $J_z$, the field above the upper cut differs radically from the field above the lower cut: 
the latter contains a ``floating'' flux rope. The key characteristic of the photospheric imprint of this coronal flux rope in the lower cut is the inverse orientation \citep{Low1995, Kuperus1974}, or bald-patch topology \citep{Titov1993}, of the horizontal field at the PIL, which points from negative $B_z$ toward positive $B_z$.  The signs of $\nablavec_h \cdot \bvec_h$ in the two cuts are opposite, so $\bvec^>(x,y,\pm z_0)$ differs dramatically between the cases.

\subsection{Future Work}

Given the key role of coronal currents in storing magnetic energy, much work to understand basic aspects of photospheric imprints in these and other ARs remains to be done. 
Observational studies of differences in active regions' $\bvec^>(x,y,0)$ --- their
morphologies and evolution --- should be investigated, along with
relationships between $\bvec^>(x,y,0)$ and flare / CME productivity.  It is known that ARs with simple bipolar structure are less flare-productive than more complex ARs (e.g., \citealt{Sammis2000}).  Does the structure of $\bvec^>(x,y,0)$ tend to be simpler in less flare-productive regions?

To relate features in imprints to coronal currents, joint analysis of coronal NLFFF models and imprints shows promise.  
A fundamental step in understanding imprints is identification of the coronal currents that produce them.  
Using the Biot-Savart law with modeled coronal currents  
$\jvec^{\rm mod}(x,y,z)$ as sources --- from, for instance, NLFFF models --- one can determine which parts of the photospheric field produced by the model, $\bvec^{\rm >, mod}(x,y,0)$, arise from which coronal currents.
We expect smaller-scale, intermittent structures in $\bvec^>(x,y,0)$ to arise from small-scale currents near the photosphere (because higher-order multipole terms decay rapidly with distance), while larger-scale features in $\bvec^>(x,y,0)$ probably arise from more coherent structures above the photosphere.
As an easier first step, directly comparing NLFFF models' $\bvec^{\rm >, mod}(x,y,0)$ with the observed $\bvec^>(x,y,0)$, as recently undertaken by \citet{Iyer2026}, can provide insights into models' shortcomings.  Model-data agreement about $\bvec^>$ is a necessary condition for the model to provide credible insights about the coronal sources of $\bvec^>(x,y,0)$.
The development of photospheric imprints as flux emerges is another important area of study.

In addition, methods to incorporate information about coronal currents inferred from $\bvec^>(x,y,0)$ into coronal field extrapolations (like NLFFF fields) should be investigated.  Because the observed photospheric field contains substantial uncertainties and is not force-free, observed values of $\bvec(x,y,0)$ have sometimes been altered (``preprocessed,'' e.g., \citealt{Wiegelmann2006}) prior to extrapolation. 
Perhaps any modifications of the observed field prior to NLFFF extrapolation could be optimized to preserve agreement between modeled and observed $\bvec^>(x,y,0)$.

\acknowledgments{{\bf Acknowledgments:} 
We thank Prof. Michael Wheatland for helpful comments on an earlier draft of this manuscript.
We thank the US taxpayers
  for providing the funding that made this research possible, and
  gratefully acknowledge support from NASA LWS 80NSSC19K0072, NASA HSR 80NSSC23K0092, NSF AGS 2302697, from joint
  NASA-NSF funding of the Coronal Global
  Evolutionary Model (CGEM) project (http://cgem.ssl.berkeley.edu/) to
  UC Berkeley through NSF award AGS1321474 and through NASA's one-year
  extension project, ``ECGEM,'' through award 80NSSC19K0622.
MGL was supported by the Office of Naval Research and by the NASA Living with a Star program under the FST topic ``Towards a Quantitative Description of the Magnetic Origins of the Corona and Inner Heliosphere,'' via Inter Agency Agreement NNH23OB19A to NRL.
NASA's SDO satellite and the HMI instrument were joint efforts by many
teams and individuals, whose efforts to produce the HMI magnetograms
that we analyzed here are greatly appreciated.
Hinode is a Japanese mission developed and launched by ISAS/JAXA,
collaborating with NAOJ as a domestic partner, and NASA and STFC (UK)
as international partners. Scientific operation of the Hinode mission
is conducted by the Hinode science team organized at ISAS/JAXA. This
team mainly consists of scientists from institutes in the partner
countries. Support for the post-launch operation is provided by JAXA
and NAOJ (Japan), STFC (UK), NASA (USA), ESA, and NSC (Norway).}
Last but not least, BTW acknowledges generous support from the Fundaci\'on Jes\'us Serra, which supported a visit to the IAC-Tenerife during which a portion of this work was completed.

\appendix

\section{Interior \& Coronal Sources of $P^<$ and $P^>$}
\label{app:PTD_sources}


As discussed in Section \ref{subsec:ptd_separation}, the poloidal potentials at the photosphere ($z = 0$), $P^>(x,y,0)$ and $P^>(x,y,0)$, are produced by currents above and below the $z = 0$ surface. 
Following \citet{Backus1986} and \citet{Backus1996}, we now explicitly describe the quantitative relationships between these potentials and their source currents in an infinite, Cartesian domain. 
The PTD representation of the magnetic field is
  \be
  \bvec=\overbrace{\nabla\times\nabla\times{P\,\hatz}}^{\bvec_\mathrm{P}}+\overbrace{\nabla\times{T\,\hatz}}^{{\bvec}_\mathrm{T}},\label{eqn:bvec:PTD}
  \ee
where $\bvec_\mathrm{P}$ and $\bvec_\mathrm{T}$ are orthogonal for suitable
boundary conditions \cite[]{Berger2018}.  The magnetic field and current are
related though Amp{\'e}re's law
\be
\nabla\times\bvec=\emfac\,\jvec.\label{eqn:Ampere}
\ee

Following \cite[\S~5.3][]{Backus1986} (or at Equation 65a in \cite{Backus1986}), the current density may also be represented in 
PTD as
\be
  \jvec=\overbrace{\nabla\times\nabla\times{\jp\,\hatz}}^{\jvec_\mathrm{P}}+\overbrace{\nabla\times{\jt\,\hatz}}^{{\jvec}_\mathrm{T}}.\label{eqn:jvec:PTD}
\ee
where $\jp$ and $\jt$ are scalar potentials for the poloidal and toroidal current components, respectively, and 
$\jvec_\mathrm{P}$ and $\jvec_\mathrm{T}$ are also orthogonal for
suitable boundary conditions.  Orthogonality with Amp{\'e}re's law 
implies the toroidal magnetic field scalar is proportional to the poloidal current scalar,
\be
T=\emfac\,\jp.
\ee
The curl of Amp{\'e}re's law~(\ref{eqn:Ampere})
implies
\be
\nabla\times\nabla\times\bvec=\emfac\,\nabla\times\jvec,
\ee
where the right hand side is the vorticity of the current.
The $\hatz$ component of this equation produces
\be
\nabla_h^2\nabla^2{P}=\emfac\nabla_h^2{\jt}
\ee
In the unbounded domain, where pesky harmonic terms can be neglected,
this initially complex fourth order partial differential equation simplifies
to a very familiar looking Poisson equation relating the poloidal magnetic
scalar to the toroidal current scalar
\be
\nabla^2{P}=\emfac{\jt},
\ee
\emph{that is valid for all $z\in(\infty,\infty)$}.\par
Suppose now that this unbounded domain is divided into two half spaces
    $z<0$ and $z>0$ at the photosphere, $z=0$. If the toroidal current scalar
    $\jt$ is known throughout all space, then Gauss decomposition at $z=0$ is
    equivalent to solving the Poisson system
\begin{subequations}
\be
\nabla^2{P^<}=\emfac\,\left\lbrace\begin{array}{cc}\jt&z<0\\ 0&z>0
\end{array}\right.\label{eqn:PLT}
\ee
\be
\nabla^2{P^>}=\emfac\,\left\lbrace\begin{array}{cc}0&z<0\\ \jt&z>0\end{array}\right..\label{eqn:PGT}
\ee
\end{subequations}
where the poloidal magnetic field
\be
\bvec_{\mathrm{P}}^{<(>)}=\nabla\times\nabla\times{P^{<(>)}}\hatz,\label{eqn:bp:PTD}
\ee
is
produced by corresponding currents below $z<0$ or above the photosphere $z>0$.
%
(In terms of the notation used in Section \ref{subsec:ptd_separation}, $P^<$ in Equation (\ref{eqn:PLT}) is identical to $P^<$ on the half-domain $z \le 0$, and $P^>$ in Equation (\ref{eqn:PGT}) is identical to $P^>$ on the half-domain $z \ge 0$.)
%
For localized sources, the boundary conditions $P^{<}\rightarrow0$ and
$P^{>}\rightarrow0$ as $|\boldsymbol{x}|\rightarrow\infty$, as well as
continuity of the potentials $P^<$ and $P^>$ across $z=0$, ensure that the
unique solution to Equations~(\ref{eqn:PLT})\--(\ref{eqn:PGT}) can be found by
standard methods (e.g., \citealt{Jackson1975}).
\par
%

The elegance of the Gauss decomposition within the framework of
  PTD is in the continuity of the mathematical representation throughout the
  domain $z\in(-\infty,\infty)$. In contrast, the representation in the
  scalar potential approach, presented Section \ref{subsec:potential_separation} above, where
  $\bvec_\mathrm{P}\sim\nabla\psi^{<(>)}$, is relegated to a domain of validity
  in one of \emph{either} of the 
  half-spaces $z\in(-\infty,0]$ or
$z\in[0,\infty)$ but not both.

\end{document}